\documentclass[journal,10pt]{IEEEtran}
 \usepackage[super]{nth} 
 \usepackage{amsthm}
 \usepackage{stfloats}
\newtheorem{example}{Example}
\usepackage{tikz}
\usetikzlibrary{shapes,arrows,positioning}
\usetikzlibrary{matrix, positioning}
\usepackage{multirow}
\usepackage{pgfplots}
\pgfplotsset{compat=1.18}
\usepackage{graphicx}
\usepackage{wrapfig}
\usepackage{placeins}
\usepackage{pdfpages}
\usepackage{attachfile2}
\usepackage{amsfonts}
\usepackage{caption}
\usepackage{enumitem}
\usepackage{epstopdf} 
\usepackage{fancyhdr}
\usepackage{amsmath,amsfonts}
\usepackage[ruled,vlined]{algorithm2e}
\usepackage{array}
\usepackage[caption=false,font=normalsize,labelfont=sf,textfont=sf]{subfig}
\usepackage{textcomp}
\usepackage{float}
\usepackage{stfloats}
\usepackage{url}
\usepackage{verbatim}
\usepackage{graphicx}
\usepackage{cite}
\usepackage{algorithmic}
\usepackage{booktabs}
\usepackage{makecell}
\usepackage{multirow}
\usetikzlibrary{shapes.geometric,arrows.meta,calc}
\definecolor{cA}{rgb}{1.00,0.00,0.00}   
\definecolor{cB}{rgb}{0.60,0.20,0.90}   
\definecolor{cC}{rgb}{0.00,0.40,0.00}   
\definecolor{cD}{rgb}{0.10,0.10,1.00}   
\definecolor{dA}{rgb}{0.13,0.13,0.55}   
\definecolor{dB}{rgb}{1.00,0.60,0.00}   
\definecolor{dC}{rgb}{0.60,0.15,0.15}   
\definecolor{dD}{rgb}{0.00,0.00,0.00}   

\newcommand{\cw}{0.62}    
\newcommand{\ch}{0.62}    
\newcommand{\ms}{0.15}    
\newcommand{\gap}{3.30}   
\newcommand{\vytop}{-3.65}

\newcommand{\Dia}[3]{\begin{scope}[shift={(#2,#3)}]
  \draw[draw=#1,thick,line join=round] (0,\ms)--(\ms,0)--(0,-\ms)--(-\ms,0)--cycle;\end{scope}}
\newcommand{\Cro}[3]{\begin{scope}[shift={(#2,#3)}]
  \draw[draw=#1,thick] (-\ms,-\ms)--(\ms,\ms); \draw[draw=#1,thick] (-\ms,\ms)--(\ms,-\ms);\end{scope}}
\newcommand{\Tri}[3]{\begin{scope}[shift={(#2,#3)}]
  \draw[draw=#1,thick,line join=round] (0,\ms)--(\ms,-\ms)--(-\ms,-\ms)--cycle;\end{scope}}
\newcommand{\Squ}[3]{\begin{scope}[shift={(#2,#3)}]
  \draw[draw=#1,thick,line join=round] (-\ms,-\ms) rectangle (\ms,\ms);\end{scope}}
\newcommand{\Rtr}[3]{\begin{scope}[shift={(#2,#3)}]
  \draw[draw=#1,thick,line join=round] (-\ms,\ms)--(\ms,\ms)--(\ms,-\ms)--cycle;\end{scope}}
\newcommand{\Sta}[3]{\node[star,star points=5,star point ratio=2.3,draw=#1,fill=#1,
  minimum size=2.9mm,inner sep=0pt] at (#2,#3) {};}

\newcommand{\codebook}[5]{%
\begin{scope}[shift={(#1,0)}]
  \foreach \r in {1,...,4}{%
    \foreach \c/\ca/\cb in {0/cA/dA, 1/cB/dB, 2/cC/dC, 3/cD/dD}{%
      \pgfmathsetmacro{\xx}{(\c+0.5)*\cw}%
      \pgfmathsetmacro{\yy}{-(\r-0.5)*\ch}%
      \draw[black!75,line width=0.4pt] (\c*\cw,-\r*\ch) rectangle ({(\c+1)*\cw},{-(\r-1)*\ch});
      \ifnum\r=#3\relax
        #5{\ca}{\xx}{\yy}%
      \else
        \ifnum\r=#4\relax
          #5{\cb}{\xx}{\yy}%
        \else
          \node[inner sep=0pt] at (\xx,\yy) {$0$};
        \fi
      \fi
    }%
  }%
  \draw[black,line width=0.9pt] (0,0) rectangle ({4*\cw},{-4*\ch});
  \node at ({2*\cw},{0.42}) {${ \mathbf{\cal{X}}_{#2}}$};
  \foreach \c in {0,...,3}{\node at ({(\c+0.5)*\cw},{-4*\ch-0.30}) {$\c$};}
\end{scope}}

\newcommand{\codeword}[8]{%
\begin{scope}[shift={(#1,0)}]
  \pgfmathsetmacro{\vx}{1.5*\cw}%
  \foreach \r in {1,...,4}{%
    \pgfmathsetmacro{\xx}{\vx+0.5*\cw}%
    \pgfmathsetmacro{\yy}{\vytop-(\r-0.5)*\ch}%
    \draw[black!75,line width=0.4pt] (\vx,{\vytop-\r*\ch}) rectangle ({\vx+\cw},{\vytop-(\r-1)*\ch});
    \ifnum\r=#3\relax
      #5{#6}{\xx}{\yy}%
    \else
      \ifnum\r=#4\relax
        #5{#7}{\xx}{\yy}%
      \else
        \node[inner sep=0pt] at (\xx,\yy) {$0$};
      \fi
    \fi
  }%
  \draw[black,line width=0.9pt] (\vx,\vytop) rectangle ({\vx+\cw},{\vytop-4*\ch});
  \node at ({\vx+0.5*\cw},{\vytop-4*\ch-0.32}) {$\mathbf{x}_{#2}$};
  \pgfmathsetmacro{\sx}{(#8+0.5)*\cw}%
  \draw[black,line width=0.6pt] ({\sx},{-4*\ch-0.30}) ellipse (0.16 and 0.24);
  \draw[-{Latex[length=1.8mm,width=1.4mm]},line width=0.6pt]
       ({\sx},{-4*\ch-0.56}) to[out=-70,in=110] ({\vx+0.5*\cw},{\vytop+0.10});
\end{scope}}
\usetikzlibrary{shapes.symbols, arrows.meta, positioning, decorations.pathreplacing, calc}

\definecolor{splitterC}{RGB}{196,181,225}   
\definecolor{pinkC}{RGB}{237,160,180}       
\definecolor{scmaC}{RGB}{238,221,151}       
\definecolor{mpaC}{RGB}{241,226,117}        
\definecolor{rxsplitC}{RGB}{162,193,235}    
\definecolor{cloudC}{RGB}{150,186,224}      
\usetikzlibrary{shapes.symbols, arrows.meta, positioning, decorations.pathreplacing, calc}

\definecolor{mpaC}{RGB}{241,226,117}        
\definecolor{extractC}{RGB}{150,200,140}    
\definecolor{lavenderC}{RGB}{196,181,225}   
\definecolor{orangeC}{RGB}{235,165,110}     
\definecolor{pinkC}{RGB}{237,160,180}       
\definecolor{cMPA}{RGB}{255,255,102}
\definecolor{cExtract}{RGB}{102,255,153}
\definecolor{cLDPC}{RGB}{255,153,255}
\definecolor{cHard}{RGB}{255,178,102}
\definecolor{cMult}{RGB}{178,178,255}
\definecolor{cComb}{RGB}{51,204,204}
\newcommand{\antenna}[2][black]{%
  \begin{scope}[shift={(#2)}, scale=0.28, line width=0.8pt, color=#1]
    \draw (0,0) -- (0,1.0);
    \draw (0,1.0) -- (-0.45,1.55);
    \draw (0,1.0) -- (0.45,1.55);
    \draw (0,1.0) -- (0,1.65);
    \filldraw (0,0) circle (0.09);
  \end{scope}
}
\begin{document}

\title{A Rate-Splitting Approach to SCMA Systems with Heterogeneous Message Passing Detector under Finite Blocklength}
\title{Heterogeneous SCMA: A Novel Code-domain RSMA Framework}
\title{Heterogeneous SCMA: Design and Optimization of a  Novel Code-domain RSMA Framework under Finite Blocklength }
\title{A  Novel Code-domain RSMA Framework based on Heterogeneous SCMA under Finite Blocklength }
\title{Design and Analysis of  SIC-Free and SIC-Based Heterogeneous RS-SCMA under Finite-Alphabet Constraints }
   \author{Ashlesha Medhi,~\IEEEmembership{Student Member,~IEEE,}%
        ~Kuntal Deka,~\IEEEmembership{Member,~IEEE,}%
        ~Zilong Liu,~\IEEEmembership{Senior Member,~IEEE,}
         Sanjeev Sharma,~\IEEEmembership{Senior Member,~IEEE},%
         ~ A. Rajesh,
         and 
         ~ Leila Musavian,~\IEEEmembership{Senior Member,~IEEE.}
        \thanks{Ashlesha Medhi, Kuntal Deka,  and A. Rajesh are Department of EEE,  Indian Institute of Technology Guwahati, Guwahati 780139, India (email: ashlesha.medhi@iitg.ac.in; kuntaldeka@iitg.ac.in, rajesha@iitg.ac.in).}%
        \thanks{Zilong Liu and Leila Musavian are with the School of Computer Science and Electrical
Engineering, University of Essex, Colchester CO4 3SQ, U.K. (e-mail:
zilong.liu@essex.ac.uk, leila.musavian@essex.ac.uk).}%
        \thanks{Sanjeev Sharma is with the Department of Electronics Engineering,  Indian Institute of Technology (BHU), India (email: sanjeev.ece@iitbhu.ac.in).}%
}

\maketitle
\begin{abstract}  
This paper proposes a rate-splitting sparse code multiple access (RS-SCMA) scheme with non-uniform message splitting, which allows flexible modulation orders across the split message components. A heterogeneous message passing algorithm (MPA) receiver is developed to support different modulation-order profiles, followed by its successive interference cancellation (SIC)-based version for efficient detection. We then derive analytical expressions for the finite-alphabet achievable rate and the bit error rate (BER) of the proposed system. For a fair comparison with conventional rate-splitting multiple access (RSMA), we also develop a multicarrier RSMA (MC-RSMA) benchmark. Its transmitter uses user-specific rotated QAM codebooks that keep the superimposed constellation points distinct, Hungarian-algorithm-based user–subcarrier matching, and power allocation that maximizes the finite-alphabet sum rate. At its receiver, the common stream is jointly detected across all subcarriers and removed by soft SIC, and the private streams are then recovered by joint maximum-likelihood (ML) detection. We further extend the framework to a low-density parity-check (LDPC)-coded RS-SCMA system, in which soft information from the decoder is used to reconstruct and cancel the common stream before the private streams are detected and decoded. Simulation results show that, at the same spectral efficiency, RS-SCMA achieves a higher sum rate and a lower BER than both conventional SCMA and MC-RSMA.
\end{abstract}
\section{Introduction}

\IEEEPARstart{R}{ate}-splitting multiple access (RSMA) has emerged as a promising multiple access framework that subsumes space-division multiple access (SDMA), power-domain non-orthogonal multiple access (NOMA), and physical layer broadcasting\cite{9831440}. In RSMA, each user's message is split into common and private messages. At the receiver, the common stream can be decoded by any user and removed through successive interference cancellation (SIC) before detection of the corresponding private stream. By allowing part of the interference to be decoded while treating the remaining interference as noise, RSMA provides a flexible mechanism for managing interference under diverse channel conditions.


Existing works on RSMA focus on Gaussian signaling at the transmitter. In contrast, real-world communication systems transmit symbols from finite-alphabet constellations. Recent
studies have investigated several practical and generalized forms of RSMA. A practical
physical-layer RSMA architecture employing finite-constellation modulation, finite-length polar
codes, and adaptive modulation and coding (AMC) was proposed in~\cite{9217326}. An SIC-free RSMA
system based on constellation-constrained optimization was investigated in~\cite{11351316}, while
a zero-SIC receiver for quadrature RSMA (Q-RSMA) was introduced in~\cite{9831449}. In Q-RSMA, the
common and private streams are mapped onto the in-phase and quadrature components, respectively,
so that both can be detected simultaneously without conventional SIC. This reduces the decoding
latency, error propagation, and receiver complexity. Furthermore, finite-constellation-constrained
sum-rate optimization for RSMA with both SIC-based and SIC-free receivers was studied
in~\cite{10485496}.

Besides finite-alphabet design, multicarrier implementation is another key
aspect of practical RSMA. In~\cite{8926413}, a multicarrier downlink
multiple-input single-output (MISO) multicast system was developed, in which
each message is split into common and private parts on individual subcarriers.
More recently, multicarrier RSMA (MC-RSMA) has been studied from the
perspectives of power allocation and user matching~\cite{9403732,11232518}.
In MC-RSMA, the achievable rate is jointly determined by the assignment of
private streams to subcarriers and by the power split between the common and
private streams.

\subsection{Motivation and Contributions}
The integration of rate-splitting (RS) and code-domain NOMA (CD-NOMA) has emerged as a promising research direction for future multiple access systems. In \cite{10038476}, it was mentioned that ``\textit{\ldots there is no effort so far on the interplay between code-domain NOMA and RSMA and this is an area of interest to see whether we can further enhance RSMA performance by bringing the code-domain dimension.}'' Motivated by this research gap, the integration of CD-NOMA with RSMA has been further studied. The integration of RSMA with other MA schemes can be better understood through the broader concept of unified multiple access (UMA), which was first introduced in \cite{10562043}. A distinctive feature of RSMA is its ability to unify MA schemes such as OMA, SDMA, PD-NOMA, and multicasting into a single framework. Owing to this unifying capability, RSMA can be further extended to the code domain, in addition to the time, frequency, power, and spatial domains.
The paper \cite{11493467} considers an SIC-based SISO hybrid RS-SCMA system. This work is the first attempt to formulate RSMA in the code-domain (SCMA), referred to as Type-I RS-SCMA. The common message is formed by combining the selected users and is modulated using M-QAM, while the private messages are transmitted using SCMA codebooks. At the receiver side, the common stream is decoded using QAM demodulator and then a soft SIC is performed to remove the common stream. After that MPA detector is used for the detection of the private streams. However, this paper has some limitations that restrict the flexibility and reliability of the RS-SCMA system. The splitting factor $\alpha$ only changes the number of symbols in the common and private streams, while the modulation order stays the same. Therefore, the sum rate is highest at $\alpha = 0.5$, and different target spectral efficiencies cannot be achieved. Moreover, a regular factor graph is used for private users, in which all users occupy the same number of resources. In addition, the common stream is sent as QAM symbols outside the SCMA factor graph and is detected by a simple QAM demodulator. This makes the common stream the weakest part of the SIC process. These limitations motivate the present work.

The main contributions of this work are as follows:
\begin{itemize}
 \item We introduce a new RS-SCMA architecture, termed Type-II RS-SCMA, in which different modulation orders are assigned to different users, resulting in a heterogeneous modulation profile. By appropriately selecting the modulation order for each user, the proposed architecture provides flexibility in controlling the allocation of resources between the common and private messages. This, in turn, enables the system to accommodate diverse spectral-efficiency requirements while offering an additional degree of flexibility in system design.
   \item We design an irregular factor graph for the proposed system. Since the common stream must be decoded by all users, it requires more protection than the private streams. Therefore, the irregular factor graph assigns more resources and higher power to the common user than to the private users.
    \item  We propose an SIC-based  MPA receiver for Type-II RS-SCMA with heterogeneous modulation order profile. The common stream is first detected and cancelled using SIC. The remaining private streams are then detected using another heterogeneous MPA on a reduced factor graph after removal of the virtual common user node.
    \item We derive the achievable sum rate expression of the proposed Type-II RS-SCMA system under finite alphabet constraint. We further propose a low-density parity-check (LDPC)-coded, SIC-based heterogeneous MPA receiver for Type-II RS-SCMA under finite-alphabet. The common stream is first detected using MPA and decoded using an LDPC decoder. After that, the decoded soft information is used to generate the soft codeword for common stream used in SIC stage. The private streams are then detected by a second MPA stage and subsequently decoded by their respective LDPC decoders.
   \item We develop a finite-alphabet MC-RSMA benchmark with user-specific constellation rotation, where the rotation angles and powers are optimized to maximize the sum rate. At the receiver, after soft SIC of the common stream, subcarrier-wise joint ML detection decodes the co-channel private streams instead of treating them as noise, thus removing the high-SNR error floor.
    
\end{itemize}
\subsection{Notations}
In this paper, regular, bold lowercase, bold uppercase, and script letters denote scalars, vectors,
matrices, and sets, respectively. We use $\mathbb{C}^{J\times1}$ for a $J\times1$ complex vector
and $\mathcal{CN}(\mu,\sigma^{2})$ for a complex Gaussian distribution with mean $\mu$ and variance
$\sigma^{2}$. The superscripts $(\cdot)^{T}$ and $(\cdot)^{H}$ denote the transpose and conjugate
transpose, $\mathbf{I}_{K}$ is the $K\times K$ identity matrix, and $\mathbb{E}[\cdot]$ is the
expectation operator. Also, $\mathrm{diag}(\mathbf{h})$ is a diagonal matrix with the entries of
$\mathbf{h}$ on its main diagonal, and $\mathcal{X}=\mathcal{X}_1\times\cdots\times\mathcal{X}_J$
is the Cartesian product of the user codebooks. The other symbols used in this paper are listed in
Table~\ref{tab:symbols}.
\begin{table}[!t]
\caption{List of Symbols and Notations.}
\label{tab:symbols}
\centering
\scriptsize
\setlength{\tabcolsep}{2pt}
\renewcommand{\arraystretch}{1.05}
\begin{tabular}{|l p{0.26\columnwidth}|l p{0.26\columnwidth}|}
\hline
\textbf{Symbol} & \textbf{Description} & \textbf{Symbol} & \textbf{Description} \\
\hline
$J$, $K$ & No.\ of users, REs/subcarriers & $L$ & No.\ of split users \\
$\mathbf{F}$, $\lambda$ & Factor matrix, overloading factor & $d_v$, $d_f$ & User/resource-node degrees \\
$d_{v_j}$, $d_{v_c}$ & Private/common stream degrees & $\mathcal{M}_k$, $\mathcal{N}_i$ & Streams on RE $k$, REs of stream $i$ \\
$\eta$, $\eta_{\mathrm{coded}}$ & Uncoded/coded spectral efficiency & $r_c$, $r_p$ & Common/private code rates \\
$\mathbf{b}_j$, $\ell_j$ & Bits of user $j$, length & $\mathbf{b}_{c,j}$, $\mathbf{b}_{p,j}$ & Common/private parts of $\mathbf{b}_j$ \\
$\mathbf{b}_c$, $\ell_c$ & Common message, length & $M_0$, $M_j$ & Modulation orders \\
$M_{\mathrm{tot}}$, $M_p$ & Superimposed/private constellation sizes & $\mathbf{h}_j$, $h_{j,k}$ & Channel vector, gain on RE $k$ \\
$\mathbf{y}_j$, $\tilde{\mathbf{y}}_j$ & Received/residual signals & $\mathbf{v}_j$, $\sigma^2$ & AWGN, noise variance \\
$\mathbf{x}_{i,m_i}$ & $m_i$th codeword of stream $i$ & $\bar{\mathbf{x}}_0$ & Soft common codeword \\
$\mathbf{x}(\mathbf{m})$, $\mathbf{x}_p(\mathbf{m}_p)$ & Superimposed/private superimposed codewords & $\mathbf{m}$, $\mathbf{m}_p$ & Codeword index vectors \\
$\mathcal{X}_j$, $\mathbf{C}_{\mathrm{mother},j}$ & Codebook, mother constellation & $\boldsymbol{\mu}_j$, $\mathbf{V}_j$, $\boldsymbol{\Delta}_j$ & Scaling, mapping, rotation matrices \\
$\phi_{j,n}$ & Rotation angle on $n$th RE of stream $j$ & $\theta$, $\varphi$ & GAM parameter, golden angle \\
$p_0$, $p_j$, $P_{\mathrm{tot}}$ & Common, private, total powers & $\beta$, $\epsilon$ & Power-splitting factor, SIC imperfection \\
$\boldsymbol{\Delta}$, $\mathcal{D}(\mathbf{m},\mathbf{n})$ & Codeword difference, its nonzero REs & $d_H(\mathbf{m},\mathbf{n})$ & Hamming distance of labels \\
$\mathbf{d}_c$, $\mathbf{d}_{p,j}$ & Common/private LDPC codewords & $L(d_{i,\ell})$ & LLR of $\ell$th coded bit of stream $i$ \\
$\rho_{j,k}$, $\mathcal{P}_k$ & Subcarrier indicator, private streams on subcarrier $k$ & $M_{\mathcal{P}_k}$ & Joint constellation size on subcarrier $k$ \\
$s^{k}_0$, $s^{k}_j$, $x^{k}_{i,m_i}$ & Symbols on subcarrier $k$ & $p^{k}_0$, $p^{k}_j$ & Powers on subcarrier $k$ \\
$\mathbf{X}_j$, $\phi_j$ & Rotated QAM codebook, its angle & $\mathcal{S}_k$ & Superimposed constellation on subcarrier $k$ \\
\hline
\end{tabular}
\end{table}
\subsection{Organization of the Paper}
Section~II reviews SCMA and conventional MC-RSMA. Section~III presents the proposed Type-II RS-SCMA
system, including its codebook design, receivers, and rate analysis. Section~IV presents the
MC-RSMA architecture under finite alphabet, with subcarrier allocation and power optimization.
Section~V discusses the simulation results, and Section~VI concludes the paper. The main symbols
are listed in Table~\ref{tab:symbols}.
\section{Preliminaries}

\subsection{SCMA}
SCMA~\cite{9369968,9000717,9817015} is a CD-NOMA technique in which multiple users are separated with judiciously designed sparse codebooks. Specifically, SCMA maps the encoded bits of each user to a multidimensional
codeword using a user-specific sparse codebook. The overloading factor is defined as $\lambda=J/K$, where $J$ is
the number of users and $K$ is the number of resource elements (REs). The system is overloaded
when $\lambda>1$. The structure of an SCMA system
is described by a sparse $K\times J$ factor matrix $\mathbf{F}$. Each row represents a resource,
and each column represents a user. If $\mathbf{F}(k,j)=1$, user $j$ transmits on resource $k$; otherwise,
it does not. Most entries of $\mathbf{F}$ are zero, so it is sparse. This sparsity determines how
users share the resources, and thus the multi-user interference and the receiver complexity. Short cycles,
especially of length four, degrade the performance and convergence of the MPA. Hence,
$\mathbf{F}$ should be designed without short cycles, which requires it to be sparse. In a regular factor matrix, the number of $1$s in every row is the same, i.e., $d_f$. Similarly, the number of $1$s in every column is fixed at $d_v$.


Each user $j$ has a $K\times M$ complex codebook
$\mathcal{X}_j=[\boldsymbol{x}_{j,1},\ldots,\boldsymbol{x}_{j,M}]$. Each column
$\boldsymbol{x}_{j,m}$ is a codeword of length $K$ with exactly $d_v$ non-zero elements, placed
according to the $j$th column of $\mathbf{F}$. Each user maps $\log_2 M$ bits to an index
$m\in\{1,\ldots,M\}$ and transmits the corresponding sparse codeword. Since all users have the same modulation order $M$, the system is called homogeneous. In contrast, when users employ different modulation orders $M_j$, $\forall j=1,2,\ldots,J$, the system is referred to as heterogeneous~\cite{10858193}.

\begin{example}
Consider $J=6$, $K=4$, $d_v=2$, and $d_f=3$, so that $\lambda=J/K=1.5$ (i.e., $150\%$ overloading).
Fig.~\ref{fig:scma_factor_graph} shows the factor matrix $\mathbf{F}$ together with the corresponding
factor graph. Here, $f_{k,j}=1$ if and only if user node (UN)~$j$ is connected to RE~$k$. For example, UN~1 occupies
REs~1 and~3, and RE~1 is shared by UNs~1, 3, and~5. The factor graph in Fig.~\ref{fig:scma_factor_graph}
determines the sparsity pattern of the codebooks in Fig.~\ref{fig:scmacb}, where the non-zero rows of
$\mathcal{X}_j$ correspond to the REs connected to UN~$j$. For instance, all codewords of $\mathcal{X}_1$
are non-zero only on REs~1 and~3. Each UN then maps its data bits to a codeword $\mathbf{x}_j\in\mathcal{X}_j$,
and the codewords of all UNs are superimposed over the $K$ REs.
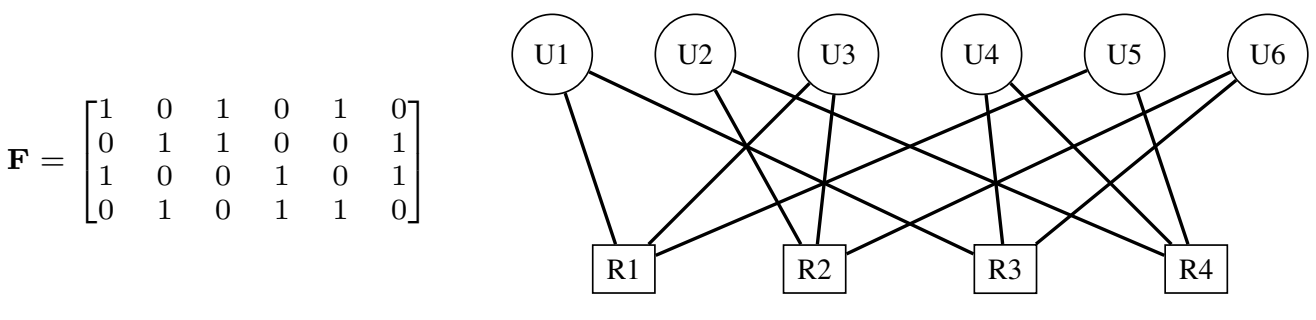
\begin{figure}[H]
    \centering
    \resizebox{0.48\textwidth}{!}{%
    \begin{tikzpicture}[
        font=\scriptsize,
        user/.style={circle, draw, minimum size=1.5mm},
        sc/.style={rectangle, draw, minimum size=2.5mm},
        line/.style={draw, thick}
    ]
    \node[anchor=east] at (0.3,0.9) {$\mathbf{F}=
        \begin{bmatrix}
        1 & 0 & 1 & 0 & 1 & 0\\
        0 & 1 & 1 & 0 & 0 & 1\\
        1 & 0 & 0 & 1 & 0 & 1\\
        0 & 1 & 0 & 1 & 1 & 0
        \end{bmatrix}$};

    \foreach \j in {1,...,6}
        \node[user] (U\j) at (\j*1.2,1.8) {U\j};
    \foreach \k/\x in {1/1.8, 2/3.4, 3/5.0, 4/6.6}
        \node[sc] (R\k) at (\x,0) {R\k};

    \foreach \k/\users in {1/{1,3,5}, 2/{2,3,6}, 3/{1,4,6}, 4/{2,4,5}}
        \foreach \j in \users
            \draw[line] (U\j) -- (R\k);
    \end{tikzpicture}%
    }
    \caption{\footnotesize Factor matrix $\mathbf{F}$ and the corresponding factor graph of an SCMA system
    with $J=6$, $K=4$, $d_v=2$, and $d_f=3$.}
    \label{fig:scma_factor_graph}
\end{figure}

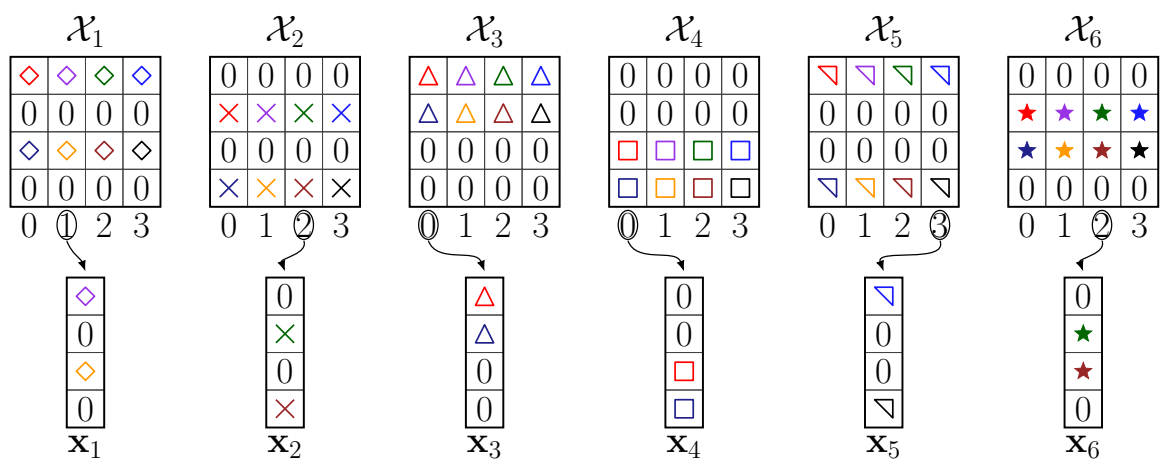
\begin{figure}[!htbp]
\centering
\scalebox{0.4}{%
\begin{tikzpicture}[font=\LARGE]
\codebook{0*\gap}{1}{1}{3}{\Dia}
\codebook{1*\gap}{2}{2}{4}{\Cro}
\codebook{2*\gap}{3}{1}{2}{\Tri}
\codebook{3*\gap}{4}{3}{4}{\Squ}
\codebook{4*\gap}{5}{1}{4}{\Rtr}
\codebook{5*\gap}{6}{2}{3}{\Sta}
\codeword{0*\gap}{1}{1}{3}{\Dia}{cB}{dB}{1}
\codeword{1*\gap}{2}{2}{4}{\Cro}{cC}{dC}{2}
\codeword{2*\gap}{3}{1}{2}{\Tri}{cA}{dA}{0}
\codeword{3*\gap}{4}{3}{4}{\Squ}{cA}{dA}{0}
\codeword{4*\gap}{5}{1}{4}{\Rtr}{cD}{dD}{3}
\codeword{5*\gap}{6}{2}{3}{\Sta}{cC}{dC}{2}
\end{tikzpicture}}
\caption{SCMA codebooks for $J=6$ and $K=4$.}
\label{fig:scmacb}
\end{figure}
\end{example}

In the downlink scenario, the base station superimposes the codewords of all $J$ users over the $K$ REs.
The received signal at user $j$ is
$
\mathbf{y}_j=\mathbf{H}_j\sum_{i=1}^{J}\mathbf{x}_i+\mathbf{v}_j,
$
where $\mathbf{H}_j=\mathrm{diag}(\mathbf{h}_j)\in\mathbb{C}^{K\times K}$,
$\mathbf{h}_j=[h_{j,1},\ldots,h_{j,K}]^{\mathrm{T}}$ contains the channel gains on the REs, and
$\mathbf{v}_j\sim\mathcal{CN}(\mathbf{0},\sigma^2\mathbf{I}_K)$ is the AWGN.
The sparsity of $\mathbf{F}$ allows the MPA to perform multi-user detection with low complexity.
In the MPA, the UNs and the RE nodes repeatedly exchange probability messages to estimate
the transmitted codewords. Since only a few users share each resource, the MPA achieves
near-maximum \textit{a posteriori} (MAP) performance with much lower complexity than exhaustive detection.
Further details can be found in~\cite{9782313}. 

\subsection{MC-RSMA}
Consider a downlink MC-RSMA system with $J$ users and $K$ subcarriers. The base station and the
users each have a single antenna. The message of each user is split into a common part and a
private part. The common parts of all users' messages are concatenated into a single common stream, whereas the private parts are mapped to the respective private streams. On subcarrier $k$, the data vector is
$\mathbf{s}^{k}=[s^{k}_{0},s^{k}_{1},\ldots,s^{k}_{J}]^T$. Here, $s^{k}_{0}$ denotes the common
stream, and $s^{k}_{j}$ denotes the private stream of user $j$. The transmitted signal on subcarrier
$k$ is
\begin{equation}
x^{k}=\sqrt{p^{k}_{0}}\,s^{k}_{0}+\sum_{j=1}^{J}\rho_{j,k}\sqrt{p^{k}_{j}}\,s^{k}_{j},
\end{equation}
where $p^{k}_{0}$ and $p^{k}_{j}$ are the common and private powers on subcarrier $k$. The
indicator $\rho_{j,k}\in\{0,1\}$ shows whether user $j$ is assigned to subcarrier $k$. The total power
constraint is
\begin{equation}
\sum_{k=1}^{K}\Big(p^{k}_{0}+\sum_{j=1}^{J}\rho_{j,k}\,p^{k}_{j}\Big)\le P_{\mathrm{tot}}.
\end{equation}
The received signal at user $j$ on subcarrier $k$ is
\begin{equation}
y_{j,k}=h_{j,k}\Big(\sqrt{p^{k}_{0}}\,s^{k}_{0}+\sum_{i=1}^{J}\rho_{i,k}\sqrt{p^{k}_{i}}\,s^{k}_{i}\Big)+v_{j,k},
\end{equation}
where $h_{j,k}\in\mathbb{C}$ is the channel coefficient and $v_{j,k}\sim\mathcal{CN}(0,\sigma^2)$
is the additive white Gaussian noise (AWGN). On each subcarrier $k$, user~$j$ first decodes the common symbol $s^{k}_{0}$, treating
all private symbols $s^{k}_{i}$ as noise. After removing $s^{k}_{0}$ by SIC, user~$j$ decodes its own
private symbol $s^{k}_{j}$ on the subcarriers with $\rho_{j,k}=1$, treating the other private symbols
$s^{k}_{i}$, $i\neq j$, as noise.

\section{System Model}

\subsection{{RS-SCMA Architecture}}
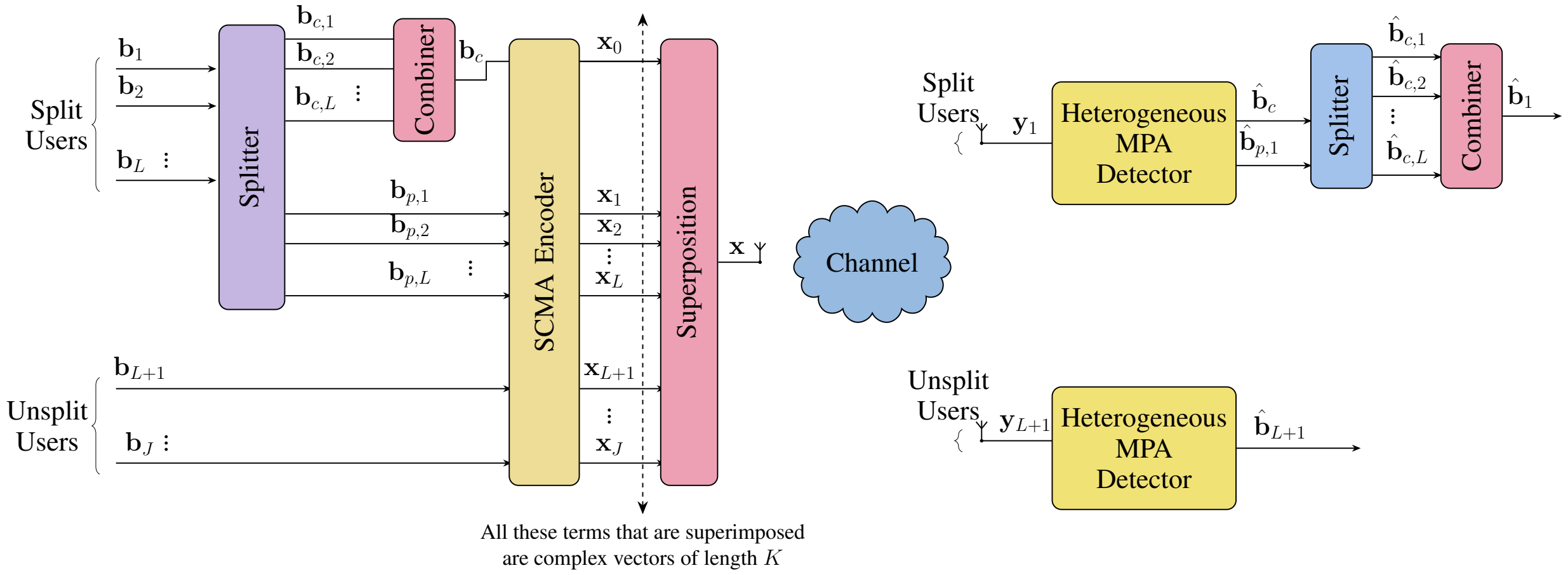
\begin{figure*}[!t]
\centering
\resizebox{0.9\textwidth}{!}{%
\begin{tikzpicture}[
    >={Stealth[length=2mm,width=1.6mm]},
    every node/.style={font=\LARGE},
    wire/.style={-{Stealth[length=2mm,width=1.6mm]}, thick},
    dwire/.style={thick, dashed},
    lbl/.style={font=\LARGE, inner sep=1pt, yshift=4pt},
    brace/.style={decorate, decoration={brace, amplitude=5pt, raise=1pt}},
]


\def\rA{11.73} \def\rB{10.88} \def\rC{10.03} \def\rD{9.18}
\def\uA{4.42} \def\uB{3.57} \def\uC{2.72}

\draw[wire] (1.215,\rA) -- (3.51,\rA) node[lbl, above, pos=0.15] {$\mathbf{b}_1$};
\draw[wire] (1.215,\rB) -- (3.51,\rB) node[lbl, above, pos=0.15] {$\mathbf{b}_2$};
\node[lbl] at (2.3625,9.605) {$\vdots$};
\draw[wire] (1.215,\rD) -- (3.51,\rD) node[lbl, above, pos=0.15] {$\mathbf{b}_L$};

\draw[wire] (1.215,\uA) -- (10.26,\uA) node[lbl, above, pos=0.06] {$\mathbf{b}_{L+1}$};
\node[lbl] at (2.3625,3.145) {$\vdots$};
\draw[wire] (1.215,\uC) -- (10.26,\uC) node[lbl, above, pos=0.06] {$\mathbf{b}_J$};

\draw[brace] (0.8775,9.18-0.255) -- (0.8775,11.73+0.255) node[midway, left=5pt, align=center] {Split\\Users};
\draw[brace] (0.8775,2.72-0.255) -- (0.8775,4.42+0.255) node[midway, left=5pt, align=center] {Unsplit\\Users};

\node[draw, thick, fill=splitterC, minimum width=1.5cm, minimum height=6.5cm, rounded corners=6pt] (splitterTX) at (4.32,9.48) {};
\node[rotate=90] at (splitterTX.center) {Splitter};

\def\cA{12.41} \def\cB{11.73} \def\cC{11.135} \def\cD{10.54}
\draw[wire] (splitterTX.east |- 0,\cA) -- (8.56,\cA) node[lbl, above, pos=0.2]{$\mathbf{b}_{c,1}$};
\draw[wire] (splitterTX.east |- 0,\cB) -- (8.56,\cB) node[lbl, above, pos=0.2, yshift=-5pt]{$\mathbf{b}_{c,2}$};
\node[lbl] at (6.75,11.135) {$\vdots$};
\draw[wire] (splitterTX.east |- 0,\cD) -- (8.56,\cD) node[lbl, above, pos=0.2]{$\mathbf{b}_{c,L}$};

\def\pA{8.415} \def\pB{7.735} \def\pC{7.14} \def\pD{6.545}
\draw[wire] (splitterTX.east |- 0,\pA) -- (10.26,\pA) node[lbl, above, pos=0.55]{$\mathbf{b}_{p,1}$};
\draw[wire] (splitterTX.east |- 0,\pB) -- (10.26,\pB) node[lbl, above, pos=0.55, yshift=-3pt]{$\mathbf{b}_{p,2}$};
\node[lbl] at (9.315,7.14) {$\vdots$};
\draw[wire] (splitterTX.east |- 0,\pD) -- (10.26,\pD) node[lbl, above, pos=0.55]{$\mathbf{b}_{p,L}$};

\node[draw, thick, fill=pinkC, minimum width=1.4cm, minimum height=2.8cm, rounded corners=6pt] (combinerTX) at (8.26,11.475) {};
\node[rotate=90] at (combinerTX.center) {Combiner};
\draw[wire] (combinerTX.east) -- ++(0.7,0) node[lbl, above, pos=0.2, xshift=6pt, yshift=6pt]{$\mathbf{b}_c$} |- (13.77,11.9);

\node[draw, thick, fill=scmaC, minimum width=1.6cm, minimum height=10.2cm, rounded corners=6pt] (scma) at (11.0,7.31) {};
\node[rotate=90] at (scma.center) {{SCMA Encoder}};

\draw[wire] (scma.east |- 0,11.9) -- (13.77,11.9) node[lbl, above, pos=0.35]{$\mathbf{x}_0$};
\draw[wire] (scma.east |- 0,\pA) -- (13.77,\pA) node[lbl, above, pos=0.35]{$\mathbf{x}_1$};
\draw[wire] (scma.east |- 0,\pB) -- (13.77,\pB) node[lbl, above, pos=0.35]{$\mathbf{x}_2$};
\node[font=\LARGE, inner sep=0pt] at (12.49,7.55) {$\vdots$};
\draw[wire] (scma.east |- 0,\pD) -- (13.77,\pD) node[lbl, above, pos=0.35]{$\mathbf{x}_L$};
\draw[wire] (scma.east |- 0,\uA) -- (13.77,\uA) node[lbl, above, pos=0.35]{$\mathbf{x}_{L+1}$};
\node[font=\LARGE, inner sep=0pt] at (12.49,3.90) {$\vdots$};
\draw[wire] (scma.east |- 0,\uC) -- (13.77,\uC) node[lbl, above, pos=0.35]{$\mathbf{x}_J$};

\draw[thick, dashed, {Stealth[length=3mm,width=2.4mm]}-{Stealth[length=3mm,width=2.4mm]}]
      (13.25,1.55) -- (13.25,13.0);
\node[align=center, font=\Large, below] at (13.25,1.45)
      {All these terms that are superimposed\\are complex vectors of length $K$};

\node[draw, thick, fill=pinkC, minimum width=1.3cm, minimum height=10.2cm, rounded corners=6pt] (super) at (14.31,7.31) {};
\node[rotate=90] at (super.center) {Superposition};
\draw[wire, -] (super.east) -- (15.93,7.31) node[lbl, above, pos=0.45]{$\mathbf{x}$};

\antenna{15.93,7.31}
\node[draw, thick, cloud, cloud puffs=15, aspect=2.3, fill=cloudC,
      minimum width=3.6cm, minimum height=2.8cm] (channel) at (18.495,7.31) {{Channel}};


\draw[brace] (20.5875,9.775) -- (20.5875,10.285) node[midway, above=10pt, xshift=-10pt, align=center]{Split\\Users};
\draw[brace] (20.5875,2.975) -- (20.5875,3.485) node[midway, above=10pt, xshift=-10pt, align=center]{Unsplit\\Users};

\antenna{20.9925,10.03}
\antenna{20.9925,3.23}

\draw[wire, -] (20.9925,10.03) -- (23.22,10.03) node[lbl, above, pos=0.45]{$\mathbf{y}_1$};
\draw[wire, -] (20.9925,3.23) -- (23.22,3.23) node[lbl, above, pos=0.45]{$\mathbf{y}_{L+1}$};

\node[draw, thick, fill=mpaC, minimum width=4.2cm, minimum height=2.8cm, rounded corners=6pt,
      align=center] (detTop) at (24.705,10.03) {{Heterogeneous}\\{MPA}\\{Detector}};
\node[draw, thick, fill=mpaC, minimum width=4.2cm, minimum height=2.8cm, rounded corners=6pt,
      align=center] (detBot) at (24.705,3.06) {{Heterogeneous}\\{MPA}\\{Detector}};

\draw[wire] (detTop.east |- 0,10.54) -- (28.54,10.54) node[lbl, above, pos=0.3, xshift=3pt]{$\hat{\mathbf{b}}_c$};
\draw[wire] (detTop.east |- 0,9.52) -- (28.54,9.52) node[lbl, above, pos=0.3]{$\hat{\mathbf{b}}_{p,1}$};
\draw[wire] (detBot.east) -- ++(2.835,0) node[lbl, above, pos=0.35]{$\hat{\mathbf{b}}_{L+1}$};

\node[draw, thick, fill=rxsplitC, minimum width=1.4cm, minimum height=3.3cm, rounded corners=6pt] (splitterRX) at (29.215,10.65) {};
\node[rotate=90] at (splitterRX.center) {{Splitter}};

\def\qA{12.0} \def\qB{11.1} \def\qC{10.2} \def\qD{9.3}
\draw[wire] (splitterRX.east |- 0,\qA) -- (31.51,\qA) node[lbl, above, pos=0.5]{$\hat{\mathbf{b}}_{c,1}$};
\draw[wire] (splitterRX.east |- 0,\qB) -- (31.51,\qB) node[lbl, above, pos=0.5, yshift=-3pt]{$\hat{\mathbf{b}}_{c,2}$};
\node[font=\LARGE, inner sep=0pt] at (30.39,10.75) {$\vdots$};
\draw[wire] (splitterRX.east |- 0,\qD) -- (31.51,\qD) node[lbl, above, pos=0.5]{$\hat{\mathbf{b}}_{c,L}$};

\node[draw, thick, fill=pinkC, minimum width=1.4cm, minimum height=3.3cm, rounded corners=6pt] (combinerRX) at (32.185,10.65) {};
\node[rotate=90] at (combinerRX.center) {Combiner};
\draw[wire] (combinerRX.east) -- ++(1.35,0) node[lbl, above, pos=0.3]{$\hat{\mathbf{b}}_1$};

\end{tikzpicture}
}
\caption{RS-SCMA system architecture: (a) transmitter structure with rate splitting and SCMA encoding, and (b) receiver structure employing heterogeneous MPA detection.}
   \label{fig:rs_scma_system}
\end{figure*}
Consider a downlink Type-II RS-SCMA system with $J$ users and $K$ REs, as shown in
Fig.~\ref{fig:rs_scma_system}. Let $\mathbf{b}_j\in\{0,1\}^{\ell_j}$ be the bit stream of user
$j$, $j=1,\ldots,J$. Each split user $j\in\{1,\ldots,L\}$ splits its bit stream as
\begin{equation}
\mathbf{b}_j=\big(\mathbf{b}_{c,j},\,\mathbf{b}_{p,j}\big),\qquad \ell_j=\ell_{c,j}+\ell_{p,j},
\end{equation}
where $\mathbf{b}_{c,j}\in\{0,1\}^{\ell_{c,j}}$ and $\mathbf{b}_{p,j}\in\{0,1\}^{\ell_{p,j}}$ are
its common and private bits. For the unsplit users $j=L+1,\ldots,J$, $\ell_{c,j}=0$ and
$\ell_{p,j}=\ell_j$. The common bits are concatenated as
$\mathbf{b}_0=(\mathbf{b}_{c,1},\ldots,\mathbf{b}_{c,L})\in\{0,1\}^{\ell_c}$, with
$\ell_c=\sum_{j=1}^{L}\ell_{c,j}$. The common and private streams use codebooks of sizes
\begin{equation}
M_0=2^{\ell_c},\qquad M_j=2^{\ell_{p,j}},
\label{eq:codebook_size}
\end{equation}
hence the $J+1$ streams may have different modulation orders. The spectral efficiency is
\begin{equation}
\eta=\frac{1}{K}\Big(\log_2 M_0+\sum_{j=1}^{J}\log_2 M_j\Big)=\frac{1}{K}\sum_{j=1}^{J}\ell_j\ \text{bits/RE}.
\label{eq:spectral_eff}
\end{equation}
By (\ref{eq:codebook_size}), a fixed modulation order for all layers restricts the possible
splits. A heterogeneous profile allows different splits for a target $\eta$. The received signal at user $j$ is
\begin{equation}
\mathbf{y}_j=\mathrm{diag}(\mathbf{h}_j)\sum_{i=0}^{J}\mathbf{x}_i+\mathbf{v}_j,
\label{eq:rsscma_receive}
\end{equation}
where $\mathbf{y}_j,\mathbf{h}_j\in\mathbb{C}^{K}$ are the received signal and the channel
vector with $h_{j,k}\sim\mathcal{CN}(0,1)$, $\mathbf{x}_i\in\mathbb{C}^{K}$ is the SCMA codeword
of stream $i$, and $\mathbf{v}_j\sim\mathcal{CN}(\mathbf{0},\sigma^{2}\mathbf{I}_{K})$ is the AWGN. Note that the power allocation for RS-SCMA is considered during the codebook design process.

For each user $j$, a non-SIC heterogeneous MPA detector is used in order to handle different
modulation orders $M_0,\ldots,M_J$. The detected common message
$\hat{\mathbf{b}}_0$ contains the common bits of all split users. These bits are extracted as
\begin{equation}
\hat{\mathbf{b}}_0=\big(\hat{\mathbf{b}}_{c,1},\ldots,\hat{\mathbf{b}}_{c,L}\big),\qquad
\hat{\mathbf{b}}_{c,j}\in\{0,1\}^{\ell_{c,j}}.
\label{eq:common_extract}
\end{equation}
The message of each user is then reconstructed as
\begin{equation}
\hat{\mathbf{b}}_j=
\begin{cases}
\big(\hat{\mathbf{b}}_{c,j},\hat{\mathbf{b}}_{p,j}\big), & j=1,\ldots,L,\\
\hat{\mathbf{b}}_{p,j}, & j=L+1,\ldots,J.
\end{cases}
\label{eq:bj_split}
\end{equation}

\begin{figure}[!htbp]
\centering
\resizebox{0.45\textwidth}{!}{%
\begin{tikzpicture}
    [
    >={Stealth[length=2mm,width=1.6mm]},
    every node/.style={font=\LARGE},
    wire/.style={-{Stealth[length=2mm,width=1.6mm]}, thick},
    lbl/.style={font=\LARGE, inner sep=1pt, yshift=4pt},
    signlbl/.style={font=\LARGE, inner sep=1pt},
    brace/.style={decorate, decoration={brace, amplitude=5pt, raise=1pt}},
]
\node[draw, thick, fill=mpaC, minimum width=4.8cm, minimum height=3.4cm, rounded corners=6pt,
      align=center] (mpaTop) at (3.4,7.4)
      {Heterogeneous\\MPA\\over $K\times(J{+}1)$\\Factor Graph};
\node[draw, thick, fill=extractC, minimum width=4.3cm, minimum height=1.8cm, rounded corners=6pt,
      align=center] (extractBox) at (10.4,7.4)
      {Extract Common\\Message};
\node[draw, thick, fill=lavenderC, minimum width=3.6cm, minimum height=1.8cm, rounded corners=6pt,
      align=center] (softDec) at (10.4,3.0) {Soft Symbol\\Generator};
\node[draw, thick, fill=orangeC, minimum width=3.9cm, minimum height=1.8cm, rounded corners=6pt,
      align=center] (multBox) at (4.2,3.0) {Multiply by\\$\mathrm{diag}(\mathbf{h}_j)$};
\node[draw, thick, fill=mpaC, minimum width=4.8cm, minimum height=3.4cm, rounded corners=6pt,
      align=center] (mpaBot) at (3.4,-0.4)
      {Heterogeneous\\MPA\\over $K\times J$\\Factor Graph};
\node[draw, thick, fill=pinkC, minimum width=1.4cm, minimum height=9.0cm, rounded corners=6pt]
      (combiner) at (15.6,3.3) {};
\node[rotate=90] at (combiner.center) {Combiner};
\node[draw, thick, circle, minimum size=0.5cm, inner sep=0pt] (sumNode) at (-1.6,3.0) {$+$};
\antenna{-1.6,7.4}
\filldraw (-1.6,7.4) circle (1.6pt);
\draw[wire] (-1.6,7.4) -- (mpaTop.west) node[lbl, above, pos=0.5]{\LARGE$\mathbf{y}_j$};
\draw[wire] (-1.6,7.4) -- (sumNode.north) node[signlbl, pos=0.9, xshift=-10pt]{\LARGE$+$};
\draw[wire] (mpaTop.east) -- (extractBox.west) node[lbl, above, pos=0.5]{\LARGE$\hat{\mathbf{b}}_0$};
\draw[wire] (extractBox.east) -- (combiner.west |- 0,7.4) node[lbl, above, pos=0.5]{\LARGE$\hat{\mathbf{b}}_{c,j}$};
\draw[thick] (5.6,5.7) -- (5.6,4.4);
\filldraw (5.6,4.4) circle (2pt);
\draw[wire] (5.6,4.4) -- (10.4,4.4) -- (softDec.north);
\node[above, font=\LARGE, inner sep=1pt, yshift=2pt] at (8.0,4.4)
      {$P(\mathbf{x}_{0,m_0}\mid\mathbf{y}_j)$};
\draw[wire] (5.6,4.4) -- (1.4,4.4) -- (1.4,1.3);
\node[above left, font=\LARGE, inner sep=1pt, yshift=2pt] at (5.5,4.4)
      {$P(\mathbf{x}_{i,m_i}\mid\mathbf{y}_j),\ i=1,\ldots,J$};
\draw[wire] (softDec.west) -- (multBox.east) node[lbl, above, pos=0.5]{\LARGE$\bar{\mathbf{x}}_0$};
\draw[wire] (multBox.west) -- (sumNode.east) node[signlbl, pos=0.85, yshift=10pt]{\LARGE$-$};
\draw[wire] (sumNode.south) -- (-1.6,-0.4) -- (mpaBot.west) node[lbl, above, pos=0.5]{\LARGE$\tilde{\mathbf{y}}_j$};
\draw[wire] (mpaBot.east) -- (combiner.west |- 0,-0.4) node[lbl, above, pos=0.5]{\LARGE$\hat{\mathbf{b}}_{p,j}$};
\draw[wire] (combiner.east) -- ++(1.2,0) node[lbl, above, pos=0.4]{\LARGE$\hat{\mathbf{b}}_j$};
\draw[brace] (-3.0,-2.3) -- (-3.0,9.3) node[midway, left=5pt, align=center] {Split\\Users};
\end{tikzpicture}
}
\caption{RS-SCMA receiver structure employing SIC-based heterogeneous MPA detection.}
\label{fig:rs_scma_system_sic_mpa}
\end{figure}
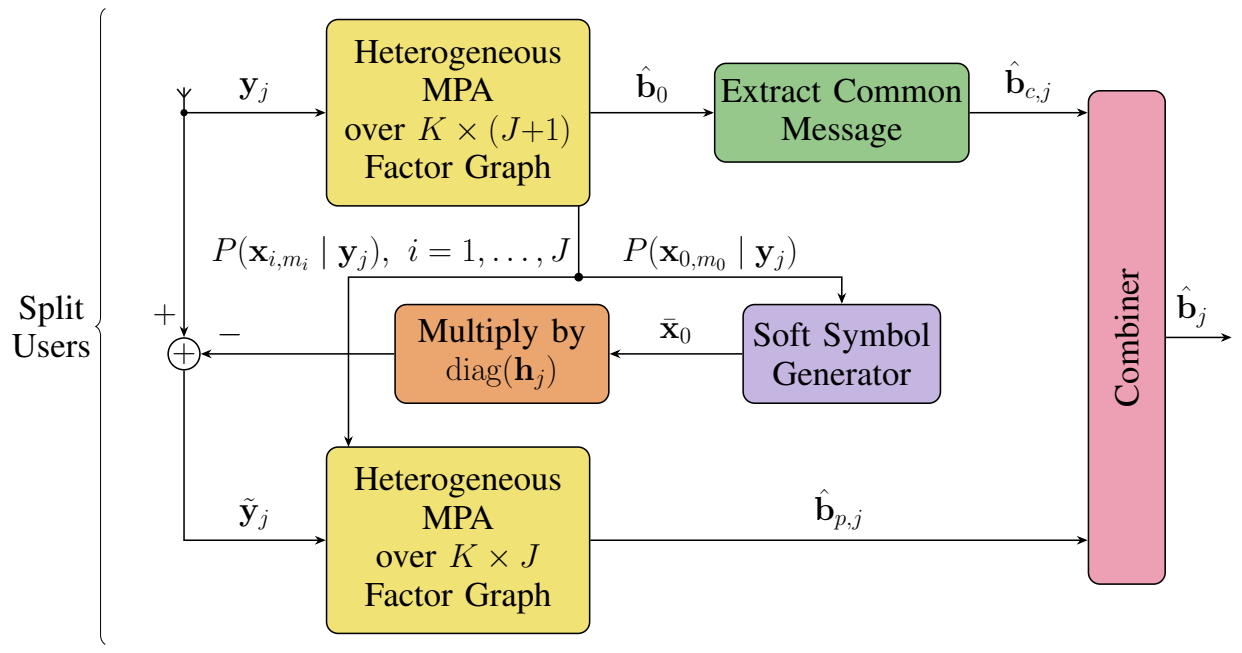

An SIC-based MPA detector is also considered, as shown in Fig.~\ref{fig:rs_scma_system_sic_mpa}.
The common stream is first detected over the $K\times(J+1)$ factor graph. Its posteriors
$P(\boldsymbol{x}_{0,m_0}\mid\mathbf{y}_j)$ give the soft common codeword
\begin{equation}
\bar{\boldsymbol{x}}_0=\sum_{m_0=1}^{M_0}P(\boldsymbol{x}_{0,m_0}\mid\mathbf{y}_j)\,\boldsymbol{x}_{0,m_0}.
\label{eq:soft_common}
\end{equation}
After the removal, we have
\begin{equation}
\tilde{\mathbf{y}}_j=\mathbf{y}_j-\mathrm{diag}(\mathbf{h}_j)\,\bar{\boldsymbol{x}}_0.
\label{eq:soft_sic}
\end{equation}
A second heterogeneous MPA then detects the private streams from $\tilde{\mathbf{y}}_j$ over the
$K\times J$ factor graph, obtained by removing the common UN. Its priors are the
first-stage posteriors $P(\boldsymbol{x}_{i,m_i}\mid\mathbf{y}_j)$, $i=1,\ldots,J$, instead of the
uniform prior $1/M_i$. The messages are then reconstructed using (\ref{eq:common_extract}) and
(\ref{eq:bj_split}).

\subsection{RS-SCMA Codebook Design}
Let $\mathcal{C}(\mathbf{F},\mathbf{M},\theta,\mathbf{p})$ be the RS-SCMA codebook for the
factor graph $\mathbf{F}$, the modulation orders $\mathbf{M}=[M_0,\ldots,M_J]$, the powers
$\mathbf{p}=[p_0,\ldots,p_J]$, and the parameter $\theta$, which sets the spiral angle of the
golden angle modulation (GAM) mother constellation in~\eqref{eq:mother_constellation}. Let $\mathrm{BER}_{\mathrm{o}}$,
$\mathrm{BER}_c$, and $\mathrm{BER}_p$ be the overall, common, and average private BER. The
codebook design is
\begin{equation}
\begin{aligned}
\min_{\mathbf{F},\,\mathbf{M},\,\theta,\,\mathbf{p}}\;\; & \mathrm{BER}_{\mathrm{o}}\\
\text{s.t.}\;\; & \textstyle\sum_{j=0}^{J}\log_2 M_j=K\eta,\\
& d_{v_0}>d_{v_j},\quad \forall j=1,\ldots,J,\\
& \textstyle\sum_{j=0}^{J}p_j=P_{\mathrm{tot}},\\
& \mathrm{BER}_c\le\mathrm{BER}_p .
\end{aligned}
\label{eq:joint_design}
\end{equation}
Problem~\eqref{eq:joint_design} is a mixed discrete--continuous problem and is hard to solve
jointly. It is therefore solved in four sequential steps. In each step, the BERs are obtained
by Monte Carlo simulation of the RS-SCMA receiver at a fixed high SNR $\gamma_{\mathrm{H}}$.

\begin{enumerate}
\item \textit{Factor-graph design:} The degree of stream $j$ is
\begin{equation}
d_{v_j}=\textstyle\sum_{k=1}^{K}\mathbf{F}(k,j).
\label{eq:user_degree}
\end{equation}
Since the common stream is decoded by all users, it is assigned more REs than the private
streams, i.e.,
\begin{equation}
d_{v_0}>d_{v_j},\qquad \forall j=1,\ldots,J.
\label{eq:common_degree}
\end{equation}
Once the degrees are fixed, $\mathbf{F}$ is chosen to minimize the number of four cycles.

\item \textit{Modulation orders:} With $\theta$ fixed to an initial value $\theta_0$ and unit
power for all streams,
\begin{equation}
\begin{aligned}
\mathbf{M}^{*}=\;&\underset{\mathbf{M}\in\mathcal{M}}{\operatorname{arg\,min}}\;\mathrm{BER}_{\mathrm{o}}\\
&\text{s.t.}\;\textstyle\sum_{j=0}^{J}\log_2 M_j=K\eta,\\
&\phantom{\text{s.t.}}\;\mathrm{BER}_c\le\mathrm{BER}_p,
\end{aligned}
\label{eq:modulation_optimization}
\end{equation}
where $\mathcal{M}$ is the set of candidate modulation orders.

\item \textit{GAM parameter:} With $\mathbf{M}^{*}$ fixed and unit power,
\begin{equation}
\theta^{*}=\underset{\theta\in\Theta}{\operatorname{arg\,min}}\;\mathrm{BER}_{\mathrm{o}},
\label{eq:theta_optimization}
\end{equation}
where $\Theta=[-\tfrac{1}{2},\tfrac{1}{2})$.

\item \textit{Power splitting:} The total power is split as
\begin{equation}
p_0=\beta P_{\mathrm{tot}},\qquad p_j=\frac{(1-\beta)P_{\mathrm{tot}}}{J},\quad \forall j=1,\ldots,J,
\label{eq:power_split}
\end{equation}
and with $\mathbf{M}^{*}$ and
$\theta^{*}$ fixed,
\begin{equation}
\begin{aligned}
\beta^{*}=\;&\underset{0<\beta<1}{\operatorname{arg\,min}}\;\mathrm{BER}_{\mathrm{o}}\\
&\text{s.t.}\;\;\mathrm{BER}_c\le\mathrm{BER}_p.
\end{aligned}
\label{eq:beta_optimization}
\end{equation}
\end{enumerate}
The final codebook is $\mathcal{C}^{*}=\mathcal{C}\big(\mathbf{F},\mathbf{M}^{*},\theta^{*},\mathbf{p}(\beta^{*})\big)$.
The construction of $\mathcal{C}$ for given $(\mathbf{F},\mathbf{M},\theta,\mathbf{p})$ is given
next.

\subsubsection{Codebook Construction}
For given $(\mathbf{F},\mathbf{M},\theta,\mathbf{p})$, the RS-SCMA codebook is built as follows. The sparse codebook of stream $j$ is $\boldsymbol{\chi}_{j}=\boldsymbol{\mu}_{j}\,\mathbf{C}_{\mathrm{mother},j}$,
where $\boldsymbol{\mu}_j=\mathbf{V}_j\boldsymbol{\Delta}_j$ is the scaling matrix and
$\mathbf{C}_{\mathrm{mother},j}$ is the mother constellation of stream $j$. The mapping matrix
$\mathbf{V}_j$ places the non-zero entries on the REs of stream $j$ given by $\mathbf{F}$, and
the rotation matrix $\boldsymbol{\Delta}_j$ rotates them on each RE. Each codeword of stream $j$ is
scaled by $\sqrt{p_j}$.

\textit{Mother constellation:} The $d_{v_j}\times M_j$ mother constellation of stream $j$ is
built by GAM~\cite{8576566} as
\begin{equation}
\mathbf{C}_{\mathrm{mother},j}=\mathrm{GAM}(M_j,\theta).
\label{eq:mother_constellation}
\end{equation}
First, $N_{p_j}=d_{v_j}M_j/2$ GAM points are generated on a spiral as
\begin{equation}
x_n=r_n\,e^{\jmath 2\pi(\varphi+\theta)n},\quad r_n=c_{\mathrm{norm}}\sqrt{n},\quad n=1,\ldots,N_{p_j},
\end{equation}
where $\varphi=\frac{\sqrt{5}-1}{2}$ is the golden angle and $c_{\mathrm{norm}}$ normalizes the
average power. These points form the first $M_j/2$ points of
$\mathbf{C}_{\mathrm{mother},j}=\{\mathbf{c}^{(j)}_1,\ldots,\mathbf{c}^{(j)}_{M_j}\}$, and the
remaining points are obtained by symmetry:
\begin{equation}
\mathbf{c}^{(j)}_{m+\frac{M_j}{2}}=-\mathbf{c}^{(j)}_m,\qquad m=1,\ldots,\tfrac{M_j}{2}.
\end{equation}
Since $d_{v_j}$ and $M_j$ differ across streams, each stream has its own mother constellation.

\textit{Rotation:} The rotation matrix of stream $j$ is
\begin{equation}
\boldsymbol{\Delta}_j=\mathrm{diag}\big(e^{\jmath\phi_{j,1}},\ldots,e^{\jmath\phi_{j,d_{v_j}}}\big),
\label{eq:rotation_matrix}
\end{equation}
where $\phi_{j,n}$ is the rotation angle on the $n$th RE of stream $j$. If RE $k$ is the $n$th RE
of stream $j$, then~\cite{8576566}
\begin{equation}
\phi_{j,n}=\frac{2\pi(c_{j,k}-1)}{M_j\,d_{f_k}},
\label{eq:rotation_angle}
\end{equation}
where $d_{f_k}$ is the number of streams on RE $k$ and $c_{j,k}$ is the order of stream $j$ among them.

\subsubsection{Analytical BER Bound}
To validate the simulated BERs, an analytical upper bound on the BER of the RS-SCMA codebook is
derived. The superimposed codeword is
$\boldsymbol{x}(\mathbf{m})=\sum_{j=0}^{J}\boldsymbol{x}_j(m_j)$,
where $\boldsymbol{x}_j(m_j)$ is the codeword chosen from $\boldsymbol{\chi}_j$ and
$\mathbf{m}=(m_0,\ldots,m_J)$ holds the codeword indices. Let $x_k(\mathbf{m})$ be the $k$th
entry of $\boldsymbol{x}(\mathbf{m})$, and let
$\mathcal{D}(\mathbf{m},\mathbf{n})=\{k\mid x_k(\mathbf{m})\neq x_k(\mathbf{n})\}$ be the set of
REs where the codewords of two index vectors $\mathbf{m}\neq\mathbf{n}$ differ.

Let $\boldsymbol{x}(\mathbf{m})$ be the transmitted codeword and $\boldsymbol{x}(\mathbf{n})$ an
erroneously detected codeword. Their difference is
\begin{equation}
\boldsymbol{\Delta}=\boldsymbol{x}(\mathbf{m})-\boldsymbol{x}(\mathbf{n})
=\sum_{i=0}^{J}\big[\boldsymbol{x}_i(m_i)-\boldsymbol{x}_i(n_i)\big],
\end{equation}
with $k$th element $\Delta(k)=x_k(\mathbf{m})-x_k(\mathbf{n})$. At the $j$th user, the pairwise
error event is
\begin{equation}
\big\|\mathbf{y}_j-\mathrm{diag}(\mathbf{h}_j)\boldsymbol{x}(\mathbf{n})\big\|^2
\le
\big\|\mathbf{y}_j-\mathrm{diag}(\mathbf{h}_j)\boldsymbol{x}(\mathbf{m})\big\|^2,
\end{equation}
which, using~\eqref{eq:rsscma_receive}, becomes
$\|\mathrm{diag}(\mathbf{h}_j)\boldsymbol{\Delta}+\mathbf{v}_j\|^2\le\|\mathbf{v}_j\|^2$.
The conditional PEP is
\begin{equation}
P_j\big(\boldsymbol{x}(\mathbf{m})\rightarrow\boldsymbol{x}(\mathbf{n})\,\big|\,\mathbf{h}_j\big)
=Q\!\left(\sqrt{\frac{\|\mathrm{diag}(\mathbf{h}_j)\boldsymbol{\Delta}\|^2}{2\sigma^2}}\right).
\end{equation}
Using $Q(x)\approx\frac{1}{12}e^{-x^2/2}+\frac{1}{4}e^{-2x^2/3}$ \cite{1210748} and averaging over independent
Rayleigh fading,
\begin{equation}
\begin{aligned}
\Pr\big(\boldsymbol{x}(\mathbf{m})\rightarrow\boldsymbol{x}(\mathbf{n})\big)
&\approx
\frac{1}{12}\prod_{k\in\mathcal{D}(\mathbf{m},\mathbf{n})}\frac{1}{1+\frac{|\Delta(k)|^2}{4\sigma^2}}\\
&\quad+\frac{1}{4}\prod_{k\in\mathcal{D}(\mathbf{m},\mathbf{n})}\frac{1}{1+\frac{|\Delta(k)|^2}{3\sigma^2}}.
\end{aligned}
\label{eq:pep}
\end{equation}
By the union bound, the overall BER is upper-bounded by
\begin{equation}
\begin{aligned}
\mathrm{BER}_{\mathrm{o}}
&\le
\frac{1}{M_{\mathrm{tot}}\log_2 M_{\mathrm{tot}}}
\sum_{\mathbf{m}}\sum_{\mathbf{n}\neq\mathbf{m}}
d_H(\mathbf{m},\mathbf{n})\\
&\qquad\times\Pr\big(\boldsymbol{x}(\mathbf{m})\rightarrow\boldsymbol{x}(\mathbf{n})\big),
\end{aligned}
\label{eq:ber}
\end{equation}
where $M_{\mathrm{tot}}=\prod_{i=0}^{J}M_i$ and $d_H(\mathbf{m},\mathbf{n})$ is the Hamming
distance between the bit labels of $\mathbf{m}$ and $\mathbf{n}$. The common and private BERs,
$\mathrm{BER}_c$ and $\mathrm{BER}_p$, are obtained by restricting the error events to the
common and private message bits. The bound in~\eqref{eq:ber} is compared with the simulated
BERs in Section~\ref{sec:results}.

\subsubsection{Unique Decodability}
For unique decodability, $\mathcal{C}^{*}$ must map different index vectors to distinct
superimposed codewords, i.e., $\boldsymbol{x}(\mathbf{m})\neq\boldsymbol{x}(\mathbf{n})$ for all
$\mathbf{m}\neq\mathbf{n}$. This holds if and only if
\begin{equation}
\mathcal{D}(\mathbf{m},\mathbf{n})\neq\emptyset,\qquad \forall\,\mathbf{m}\neq\mathbf{n},
\label{eq:ud_condition}
\end{equation}
where, on RE $k$,
\begin{equation}
x_k(\mathbf{m})-x_k(\mathbf{n})
=\sum_{j:\,m_j\neq n_j}\big[\boldsymbol{x}_j(m_j)-\boldsymbol{x}_j(n_j)\big]_k .
\label{eq:re_difference}
\end{equation}
The entries in each row of $\mathbf{C}_{\mathrm{mother},j}$ are distinct, since the radii
$r_n=c_{\mathrm{norm}}\sqrt{n}$ are distinct. Hence, if an RE carries only one stream whose
symbol differs between $\mathbf{m}$ and $\mathbf{n}$, the two codewords differ on that RE. On REs
shared by several differing streams, \eqref{eq:ud_condition} is met by the higher common power
$p_0=\beta^{*}P_{\mathrm{tot}}>p_j$, the distinct mother constellations for different $M_j$,
and the distinct radii of the rows.

\subsubsection{Distance Properties}
The MED and MPD are not used in the design. They are defined here only to compare the optimized
RS-SCMA codebook with conventional SCMA codebooks. The MED is the smallest distance between two
different superimposed codewords,
\begin{equation}
\mathrm{MED}=\min_{\mathbf{m}\neq\mathbf{n}}\big\|\boldsymbol{x}(\mathbf{m})-\boldsymbol{x}(\mathbf{n})\big\|,
\label{eq:med}
\end{equation}
and the MPD is
\begin{equation}
\mathrm{MPD}=\min_{\mathbf{m}\neq\mathbf{n}}\prod_{k\in\mathcal{D}(\mathbf{m},\mathbf{n})}\big|x_k(\mathbf{m})-x_k(\mathbf{n})\big|.
\label{eq:mpd}
\end{equation}
A larger MED improves the performance in AWGN channels, while a larger MPD improves the
performance in fading channels, as seen from the product form of~\eqref{eq:pep}.
\subsection{Achievable Rate Analysis}

The achievable rate is the highest rate at which data can be sent reliably. It depends on the
codebooks used. Rate analysis of proposed RS-SCMA considers finite alphabet. Since the rate also
depends on the receiver, so it is evaluated for both the non-SIC and the SIC-based
heterogeneous MPA receivers.
\paragraph{Non-SIC Heterogeneous MPA Receiver}
The MPA jointly detects all streams without SIC. The transmitted superimposed codeword
$\mathbf{x}(\mathbf{m})$ and a candidate $\hat{\mathbf{x}}(\mathbf{n})$ belong to the
superimposed constellation $\mathcal{X}$ of size $M_{\mathrm{tot}}=\prod_{j=0}^{J}M_j$.
The achievable sum rate is~\cite{10242334}
{\begin{equation}
\begin{aligned}
R_{\mathrm{j}}^{\mathrm{NS}}
&= \log_2 M_{\mathrm{tot}}
- \tfrac{1}{M_{\mathrm{tot}}}
\sum_{\boldsymbol{x}(\mathbf{m}) \in \mathcal{X}}
\mathbb{E}_{\mathbf{v}}
\bigg[
\log_2
\sum_{\hat{\boldsymbol{x}}(\mathbf{m}) \in \mathcal{X}}
\exp\!\bigg(
\\
&\quad
-\frac{1}{\sigma^2}
\Big\|
\mathrm{diag}(\mathbf{h}_j)\big(\boldsymbol{x}(\mathbf{m})-\hat{\boldsymbol{x}}(\mathbf{n})\big)+\mathbf{v}_j
\Big\|^2
\bigg)
\bigg].
\end{aligned}
\label{rate_non_sic}
\end{equation}}

\paragraph{SIC-Based Heterogeneous MPA Receiver}
The common stream ($j=0$) is decoded first and removed, and the private streams are then decoded
from the residual signal. In the first stage, the common stream is detected by the same joint MPA
over the $K\times(J+1)$ factor graph as in the non-SIC receiver. Hence, its rate is unchanged,
{\begin{equation}
R_0^{\mathrm{SIC}}
=
R_0^{\mathrm{NS}}.
\label{rate_common_sic}
\end{equation}}
Let $\boldsymbol{x}_p(\mathbf{m}_p)=\sum_{j=1}^{J}\boldsymbol{x}_{j}(m_j)$, with
$\mathbf{m}_p=(m_1,\ldots,m_J)$, be the transmitted private codeword, and
$\hat{\boldsymbol{x}}_p(\mathbf{n}_p)$ a candidate. Both belong to the private constellation
$\mathcal{X}_p$ of size $M_p=\prod_{j=1}^{J}M_j$. With the SIC imperfection factor
$\epsilon\in[0,1]$, the achievable private sum rate is
\begin{equation}
\begin{aligned}
&R_p^{\mathrm{SIC}}
= \log_2 M_p
- \tfrac{1}{M_p}
\sum_{\boldsymbol{x}_p(\mathbf{m}_p) \in \mathcal{X}_p}
\mathbb{E}_{\mathbf{v}}
\bigg[
\log_2
\sum_{\hat{\boldsymbol{x}}_p(\mathbf{n}_p) \in \mathcal{X}_p}
\exp\!\bigg(
\\
&\quad
-\frac{1}{\sigma^2}
\Big\|
\mathrm{diag}(\mathbf{h}_j)\big(\boldsymbol{x}_p(\mathbf{m}_p)-\hat{\boldsymbol{x}}_p(\mathbf{n}_p)
+\sqrt{\epsilon}\,\boldsymbol{x}_0\big)
+\mathbf{v}_j
\Big\|^2
\bigg)
\bigg],
\end{aligned}
\label{rate_private_sic}
\end{equation}
where $\boldsymbol{x}_0\in\mathcal{X}_0$ is the residual common codeword. The sum rate of the
SIC-based receiver is \begin{equation}
R_{\mathrm{sum}}^{\mathrm{SIC}}=R_0^{\mathrm{SIC}}+R_p^{\mathrm{SIC}}.
\label{eq:sum_rate_sic}
\end{equation}
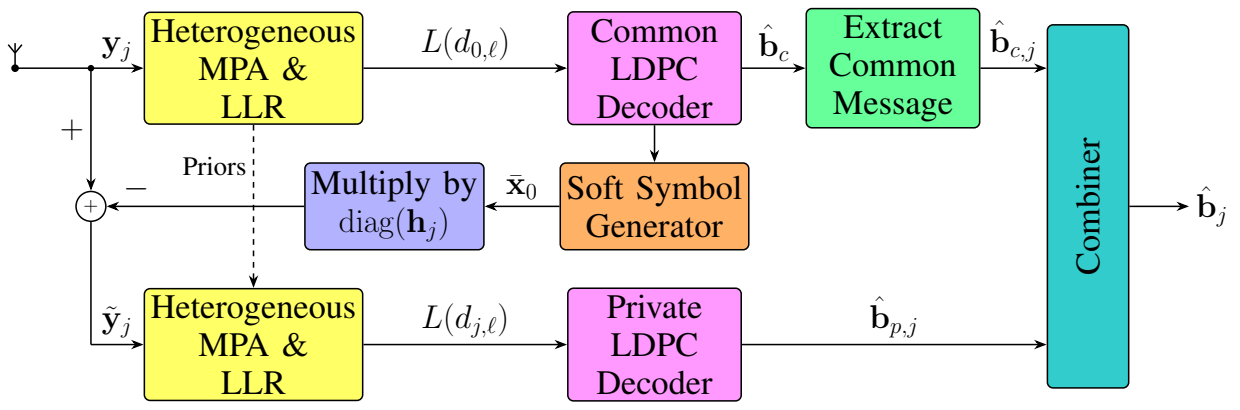
\begin{figure}[!htbp]
\centering
\resizebox{0.45\textwidth}{!}{%
\begin{tikzpicture}[
    box/.style={rectangle, draw, rounded corners=3pt, minimum width=3.2cm,
                minimum height=1.6cm, align=center, font=\LARGE, thick},
    sumnode/.style={circle, draw, minimum size=0.55cm, inner sep=0pt, thick},
    >=Stealth, thick
]
\node[box, fill=cMPA]     (s1mpa)   at (0,4.5)    {Heterogeneous\\MPA \&\\LLR};
\node[box, fill=cLDPC]    (cldpc)   at (7.4,4.5)  {Common\\LDPC\\Decoder};
\node[box, fill=cExtract] (extract) at (11.8,4.5) {Extract\\Common\\Message};
\node[sumnode]         (sum)  at (-3.0,1.95) {$+$};
\node[box, fill=cMult] (mult) at (2.6,1.95)  {Multiply by\\$\mathrm{diag}(\mathbf{h}_j)$};
\node[box, fill=cHard] (soft) at (7.4,1.95)  {Soft Symbol\\Generator};
\node[box, fill=cMPA]  (s2mpa) at (0,-0.6)   {Heterogeneous\\MPA \&\\LLR};
\node[box, fill=cLDPC] (pldpc) at (7.4,-0.6) {Private\\LDPC\\Decoder};
\node[box, fill=cComb, minimum width=1.5cm, minimum height=6.7cm]
    (comb) at (15.4,1.95) {\rotatebox{90}{Combiner}};
\antenna{-4.4,4.5}
\filldraw (-4.4,4.5) circle (1.6pt);
\coordinate (yjtap) at (-3.0,4.5);
\draw[thick]    (-4.4,4.5) -- (yjtap);
\draw[->,thick] (yjtap) -- node[above]{\LARGE$\mathbf{y}_j$} (s1mpa.west);
\filldraw       (yjtap) circle (1.3pt);
\draw[->,thick] (s1mpa.east)   -- node[above]{\LARGE$L(d_{0,\ell})$} (cldpc.west);
\draw[->,thick] (cldpc.east)   -- node[above]{\LARGE$\hat{\mathbf{b}}_c$} (extract.west);
\draw[->,thick] (extract.east) -- node[above]{\LARGE$\hat{\mathbf{b}}_{c,j}$} (comb.west |- extract.east);
\draw[->,thick] (cldpc.south) -- (soft.north);
\draw[->,thick] (soft.west) -- node[above]{\LARGE$\bar{\mathbf{x}}_0$} (mult.east);
\draw[->,thick] (mult.west) -- node[above,pos=0.85]{\LARGE$-$} (sum.east);
\draw[->,thick] (yjtap) -- node[left]{\LARGE$+$} (sum.north);
\draw[thick]    (sum.south) -- (sum.south |- s2mpa.west);
\draw[->,thick] (sum.south |- s2mpa.west) -- node[above]{\LARGE$\tilde{\mathbf{y}}_j$} (s2mpa.west);
\draw[->,thick,dashed] (s1mpa.south) -- node[left,pos=0.25]{\Large Priors} (s2mpa.north);
\draw[->,thick] (s2mpa.east) -- node[above]{\LARGE$L(d_{j,\ell})$} (pldpc.west);
\draw[->,thick] (pldpc.east) -- node[above]{\LARGE$\hat{\mathbf{b}}_{p,j}$} (comb.west |- pldpc.east);
\draw[->,thick] (comb.east) -- ++(1.1,0) node[right]{\LARGE$\hat{\mathbf{b}}_j$};
\end{tikzpicture}
}
\caption{SIC-based MPA receiver architecture of LDPC-coded RS-SCMA.}
\label{fig:sic_ldpc_mpa_receiver}
\end{figure}
\subsection{LDPC-Coded RS-SCMA System}

The common message $\mathbf{b}_0$ and the private messages $\mathbf{b}_{p,j}$, $\forall j=1,\ldots,J$,
are encoded by separate binary LDPC encoders. Suppose, the resulting codewords are $\mathbf{d}_c$ and
$\{d_{p,j}\}_{j=1}^{J}$, respectively. Two receivers are considered, namely non-SIC and SIC Heterogeneous MPA receivers.

\paragraph{Non-SIC Heterogeneous MPA Receiver}
The heterogeneous MPA jointly detects all streams over the $K\times(J+1)$ factor graph. Its
posteriors are converted into bit-wise LLRs as
\begin{equation}
L(d_{i,\ell})
=
\ln
\frac{\sum_{m_i:\,b_\ell(\boldsymbol{x}_{i,m_i})=0}P(\boldsymbol{x}_{i,m_i}\mid\mathbf{y}_j)}
     {\sum_{m_i:\,b_\ell(\boldsymbol{x}_{i,m_i})=1}P(\boldsymbol{x}_{i,m_i}\mid\mathbf{y}_j)},
\label{eq:stream_llr}
\end{equation}
where $i=0$ is the common stream and $i=1,\ldots,J$ are the private streams. Here, $d_{i,\ell}$
is the $\ell$th coded bit of stream $i$, with $\ell=1,\ldots,\log_2 M_i$. The term
$b_\ell(\boldsymbol{x}_{i,m_i})$ is the $\ell$th bit in the label of $\boldsymbol{x}_{i,m_i}$.
The bitwise LLRs are passed to the common and private LDPC decoders. The estimated relevent bits are combined to produce $\hat{\mathbf{b}}_0$ and
$\hat{\mathbf{b}}_{p,j}$. The user messages are then recovered using (\ref{eq:common_extract})
and (\ref{eq:bj_split}).

\paragraph{SIC-Based Heterogeneous MPA Receiver}
The SIC-based receiver is shown in Fig.~\ref{fig:sic_ldpc_mpa_receiver}. First, the Stage-1 MPA
computes the posteriors over the $K\times(J+1)$ factor graph. The common-stream LLRs are found
from (\ref{eq:stream_llr}) with $i=0$. The common LDPC decoder uses them to obtain
$\hat{\mathbf{b}}_c$. The \textit{a posteriori} LLRs of this decoder are converted into codeword probabilities, which replace $P(\boldsymbol{x}_{0,m_0}\mid\mathbf{y}_j)$ in
(\ref{eq:soft_common}). The common stream is then removed using (\ref{eq:soft_sic}). Next, the
Stage-2 MPA processes the residual signal $\tilde{\mathbf{y}}_j$ over the $K\times J$ factor
graph. It uses the Stage-1 posteriors of the private streams as priors. Its posteriors are
converted into LLRs using (\ref{eq:stream_llr}), with $\mathbf{y}_j$ replaced by
$\tilde{\mathbf{y}}_j$. The private LDPC decoders use these LLRs to obtain
$\hat{\mathbf{b}}_{p,j}$. Finally, the user messages are recovered as in the non-SIC receiver. The complete receiver procedure, for both the uncoded and the LDPC-coded cases, is summarized
in Algorithm~\ref{alg:rs_scma_receiver}.
\begin{algorithm}[!htbp]
\DontPrintSemicolon
\SetAlgoLined
\SetKwInOut{Input}{Input}
\SetKwInOut{Output}{Output}
\caption{\small Heterogeneous MPA-Based Type-II RS-SCMA Receiver}
\label{alg:rs_scma_receiver}
\small
\Input{$\mathbf{y}_j$, $\mathbf{h}_j$, $\sigma^2$, $\mathbf{F}$, $\{\mathbf{X}_i\}_{i=0}^{J}$}
\Output{$\{\hat{\mathbf{b}}_j\}_{j=1}^{J}$}
\BlankLine
Compute $P(\boldsymbol{x}_{i,m_i}\mid\mathbf{y}_j)$, $i=0,\ldots,J$, by MPA over the
$K\times(J+1)$ graph\;
\eIf{non-SIC}{
    \eIf{uncoded}{
        Detect $\hat{m}_i=\arg\max_{m_i}P(\boldsymbol{x}_{i,m_i}\mid\mathbf{y}_j)$,
        $i=0,\ldots,J$\;
    }{
        Compute the LLRs via (\ref{eq:stream_llr}), $i=0,\ldots,J$\;
        LDPC decode $\hat{\mathbf{b}}_c$ and $\hat{\mathbf{b}}_{p,j}$\;
    }
}{
    \eIf{uncoded}{
        Detect $\hat{m}_0=\arg\max_{m_0}P(\boldsymbol{x}_{0,m_0}\mid\mathbf{y}_j)$\;
        Compute $\bar{\boldsymbol{x}}_0$ via (\ref{eq:soft_common})\;
    }{
        Compute the LLRs via (\ref{eq:stream_llr}) with $i=0$, and LDPC decode
        $\hat{\mathbf{b}}_c$\;
        Compute $\bar{\boldsymbol{x}}_0$ via (\ref{eq:soft_common}) using the decoder
        a posteriori probabilities\;
    }
    Compute $\tilde{\mathbf{y}}_j$ via (\ref{eq:soft_sic})\;
    Compute $P(\boldsymbol{x}_{i,m_i}\mid\tilde{\mathbf{y}}_j)$, $i=1,\ldots,J$, by MPA over
    the $K\times J$ graph with Stage-1 priors\;
    \eIf{uncoded}{
        Detect $\hat{m}_j=\arg\max_{m_j}P(\boldsymbol{x}_{j,m_j}\mid\tilde{\mathbf{y}}_j)$,
        $j=1,\ldots,J$\;
    }{
        Compute the LLRs via (\ref{eq:stream_llr}) with $\tilde{\mathbf{y}}_j$,
        $i=1,\ldots,J$\;
        LDPC decode $\hat{\mathbf{b}}_{p,j}$\;
    }
}
Recover $\hat{\mathbf{b}}_j$ via (\ref{eq:common_extract}) and (\ref{eq:bj_split})\;
\end{algorithm}
\vspace{-0.2in}
\section{MC-RSMA under Finite Alphabet}
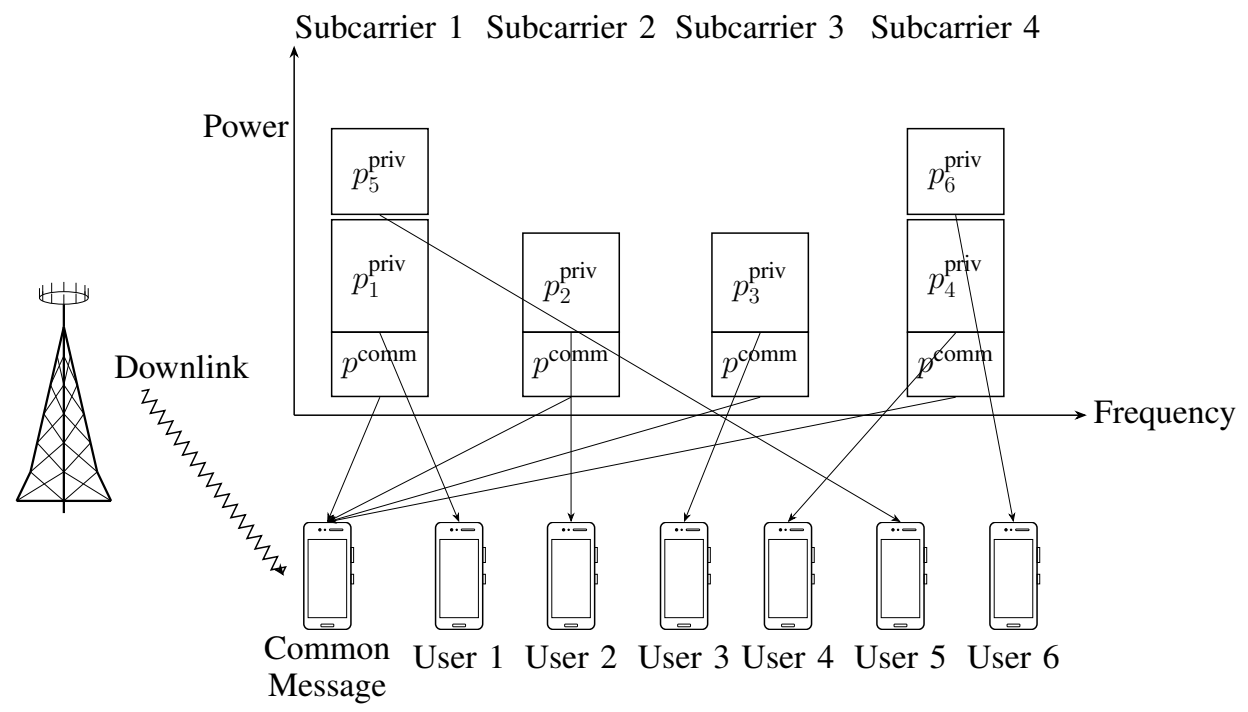
\begin{figure}[!htbp]
\centering
\resizebox{0.45\textwidth}{!}{%
\begin{tikzpicture}[
    >=Stealth,
    priv/.style={draw, thick, minimum width=1.8cm, inner sep=1pt, font=\LARGE},
    comm/.style={draw, thick, minimum width=1.8cm, minimum height=1.2cm, inner sep=1pt, font=\LARGE},
    phone/.style={
      rectangle, minimum width=1.0cm, minimum height=2.0cm, inner sep=0pt,
      path picture={
        \draw[thick, rounded corners=3pt]
          ($(path picture bounding box.south)+(-0.44,0)$) rectangle
          ($(path picture bounding box.north)+(0.44,0)$);
        \draw[fill=black] ($(path picture bounding box.north)+(-0.14,-0.14)$) circle (0.022);
        \draw[fill=black] ($(path picture bounding box.north)+(-0.045,-0.14)$) circle (0.013);
        \draw[rounded corners=0.4pt]
          ($(path picture bounding box.north)+(0.04,-0.155)$) rectangle ++(0.25,0.03);
        \draw[thin]
          ($(path picture bounding box.south)+(-0.36,0.22)$) rectangle
          ($(path picture bounding box.north)+(0.36,-0.32)$);
        \draw[rounded corners=0.4pt]
          ($(path picture bounding box.south)+(-0.12,0.05)$) rectangle ++(0.24,0.05);
        \draw ($(path picture bounding box.east)+(-0.06,0.25)$) rectangle ++(0.05,0.28);
        \draw ($(path picture bounding box.east)+(-0.06,-0.15)$) rectangle ++(0.05,0.18);
      }
    }
]

\def\xa{1.6}
\def\xb{5.17}
\def\xc{8.7}
\def\xd{12.35}

\draw[->, thick] (0,0) -- (0,6.9);
\node[font=\LARGE] at (-0.9,5.4) {Power};
\draw[->, thick] (0,0) -- (14.8,0) node[right, font=\LARGE] {Frequency};

\node[font=\LARGE] at (\xa,7.25) {Subcarrier 1};
\node[font=\LARGE] at (\xb,7.25) {Subcarrier 2};
\node[font=\LARGE] at (\xc,7.25) {Subcarrier 3};
\node[font=\LARGE] at (\xd,7.25) {Subcarrier 4};

\node[comm] (c1) at (\xa,0.95) {$p^{\text{comm}}$};
\node[priv, minimum height=2.1cm] (p1) at (\xa,2.6) {$p_1^{\text{priv}}$};
\node[priv, minimum height=1.6cm] (p5) at (\xa,4.55) {$p_5^{\text{priv}}$};

\node[comm] (c2) at (\xb,0.95) {$p^{\text{comm}}$};
\node[priv, minimum height=1.85cm] (p2) at (\xb,2.475) {$p_2^{\text{priv}}$};

\node[comm] (c3) at (\xc,0.95) {$p^{\text{comm}}$};
\node[priv, minimum height=1.85cm] (p3) at (\xc,2.475) {$p_3^{\text{priv}}$};

\node[comm] (c4) at (\xd,0.95) {$p^{\text{comm}}$};
\node[priv, minimum height=2.1cm] (p4) at (\xd,2.6) {$p_4^{\text{priv}}$};
\node[priv, minimum height=1.6cm] (p6) at (\xd,4.55) {$p_6^{\text{priv}}$};

\def\ry{-3.0}
\node[phone] (common) at (0.62,\ry) {};
\node[phone] (u1) at (3.08,\ry) {};
\node[phone] (u2) at (5.17,\ry) {};
\node[phone] (u3) at (7.28,\ry) {};
\node[phone] (u4) at (9.22,\ry) {};
\node[phone] (u5) at (11.32,\ry) {};
\node[phone] (u6) at (13.43,\ry) {};

\node[font=\LARGE, align=center] at (0.62,\ry-1.75) {Common\\Message};
\node[font=\LARGE] at (3.08,\ry-1.5) {User 1};
\node[font=\LARGE] at (5.17,\ry-1.5) {User 2};
\node[font=\LARGE] at (7.28,\ry-1.5) {User 3};
\node[font=\LARGE] at (9.22,\ry-1.5) {User 4};
\node[font=\LARGE] at (11.32,\ry-1.5) {User 5};
\node[font=\LARGE] at (13.43,\ry-1.5) {User 6};

\draw[->] (c1.south) -- (common.north);
\draw[->] (c2.south) -- (common.north);
\draw[->] (c3.south) -- (common.north);
\draw[->] (c4.south) -- (common.north);

\draw[->] (p1.south) -- (u1.north);
\draw[->] (p2.south) -- (u2.north);
\draw[->] (p3.south) -- (u3.north);
\draw[->] (p4.south) -- (u4.north);
\draw[->] (p5.south) -- (u5.north);
\draw[->] (p6.south) -- (u6.north);

\begin{scope}[shift={(-4.3,-1.6)}, scale=0.65]
\def\H{5.0}
\def\W{1.15}

\def\poleH{0.65}
\def\erx{0.7}
\def\ery{0.18}
\draw[very thick] (0,\H) -- (0,\H+\poleH);
\draw (0,{\H+\poleH+\ery}) ellipse ({\erx} and {\ery});
\foreach \ang in {0,30,...,330}{
  \draw ({\erx*cos(\ang)}, {\H+\poleH+\ery+\ery*sin(\ang)}) -- ++(0,0.28);
}

\draw[very thick] (0,5.000) -- (-0.958,0.833);
\draw[very thick] (0,5.000) -- (0.958,0.833);
\draw[very thick] (-0.958,0.833) -- (-1.357,0);
\draw[very thick] (0.958,0.833) -- (1.357,0);
\draw[very thick] (0,5.000) -- (0,-0.35);

\draw (-0.000,5.000) -- (0,4.167);
\draw (0,5.000) -- (-0.192,4.167);
\draw (0.000,5.000) -- (0,4.167);
\draw (0,5.000) -- (0.192,4.167);
\draw (-0.192,4.167) -- (0,3.333);
\draw (0,4.167) -- (-0.383,3.333);
\draw (0.192,4.167) -- (0,3.333);
\draw (0,4.167) -- (0.383,3.333);
\draw (-0.383,3.333) -- (0,2.500);
\draw (0,3.333) -- (-0.575,2.500);
\draw (0.383,3.333) -- (0,2.500);
\draw (0,3.333) -- (0.575,2.500);
\draw (-0.575,2.500) -- (0,1.667);
\draw (0,2.500) -- (-0.767,1.667);
\draw (0.575,2.500) -- (0,1.667);
\draw (0,2.500) -- (0.767,1.667);
\draw (-0.767,1.667) -- (0,0.833);
\draw (0,1.667) -- (-0.958,0.833);
\draw (0.767,1.667) -- (0,0.833);
\draw (0,1.667) -- (0.958,0.833);
\draw (-0.958,0.833) -- (0,0.000);
\draw (0,0.833) -- (-1.357,0.000);
\draw (0.958,0.833) -- (0,0.000);
\draw (0,0.833) -- (1.357,0.000);

\draw[very thick] (-1.357,0) -- (0,0) -- (1.357,0);
\end{scope}

\draw[thick, ->, decorate, decoration={zigzag, segment length=7pt, amplitude=3.5pt}]
    (-2.85,0.45) -- ($(common.west) + (-0.3,0)$);
\node[font=\LARGE] at (-2.1,0.9) {Downlink};

\end{tikzpicture}
}
\caption{ User-subcarrier allocation in MC-RSMA system}
\label{fig:mcrsma}
\end{figure}

An MC-RSMA system with $J$ users and $K$ subcarriers is considered, as shown in
Fig.~\ref{fig:mcrsma}. The common and private streams are sent over these subcarriers.
Conventional MC-RSMA usually assumes Gaussian signaling and treats the other private streams as noise. In contrast, the considered MC-RSMA system considers finite-alphabet inputs. It is also overloaded with $J>K$, so several private streams share each subcarrier. The same common
symbol is repeated on all $K$ subcarriers to obtain frequency diversity, while the private
streams are assigned to the subcarriers by Hungarian matching. User-wise rotated QAM codebooks
keep all superimposed points distinct, and the powers are allocated to maximize the
finite-alphabet sum rate. At the receiver, the common stream is detected jointly over all
subcarriers and removed by soft SIC, and the private streams are recovered by joint ML
detection.
\vspace{-0.1in}
\subsection{Subcarrier Matching with Rotated Codebooks and Joint ML Detection}
In the considered MC-RSMA system, the common and private streams are assigned to the subcarriers in
different ways. Since the common stream must be decoded by all users, it is sent on all
subcarriers, i.e.,
\begin{equation}
\rho_{0,k}=1,\qquad \forall k=1,\ldots,K.
\label{eq:common_alloc}
\end{equation}
Thus, the same common symbol is repeated on every subcarrier, i.e.,
$x^{k}_{0,m_0}=x_{0,m_0}\in\mathbf{X}_0$, $\forall k$. The
private streams are assigned by the Hungarian algorithm~\cite{9201435}. It maximizes the total
channel gain as
\begin{equation}
\begin{aligned}
\rho_{j,k}^{*}=\;&\underset{\{\rho_{j,k}\}}{\arg\max}\;\sum_{j=1}^{J}\sum_{k=1}^{K}\rho_{j,k}|h_{j,k}|^2\\
&\text{s.t.}\;\textstyle\sum_{k}\rho_{j,k}=1\;\forall j,\\
&\phantom{\text{s.t.}}\;\textstyle\sum_{j}\rho_{j,k}\ge1\;\forall k.
\end{aligned}
\label{eq:hungarian}
\end{equation}
The Hungarian algorithm needs a square cost matrix. Hence, when $J>K$, $J-K$ dummy subcarriers
with zero gain are added. The streams placed on dummy subcarriers are then moved to their
strongest subcarrier, $k^{*}=\arg\max_{k}|h_{j,k}|^2$. The set of private streams assigned to
subcarrier $k$ is
\begin{equation}
\mathcal{P}_k=\{j\in\{1,\ldots,J\}\mid\rho_{j,k}=1\},
\label{eq:private_set}
\end{equation}
which excludes the common stream $j=0$.

The codebook of stream $j$ is a rotated QAM constellation,
$\mathbf{X}_j=\{e^{\jmath\phi_j}q\mid q\in\mathcal{Q}_{M_j}\}$, $\forall j=0,1,\ldots,J$. Here, $\phi_j$ is the rotation
angle, and $\mathcal{Q}_{M_j}$ is the unit-power $M_j$-QAM constellation. Let
$\mathbf{p}=[p_0,p_1,\ldots,p_J]^{T}\in\mathbb{R}^{J+1}$ be the power vector. Using
(\ref{eq:common_alloc}), the received signal at user $j$ on subcarrier $k$ is
\begin{equation}
y_{j,k}=h_{j,k}\Big(\sqrt{p_0}\,x^{k}_{0,m_0}+\sum_{i\in\mathcal{P}_k}\sqrt{p_i}\,x^{k}_{i,m_i}\Big)+v_{j,k},
\label{eq:rsma_receiver}
\end{equation}
where $i\in\mathcal{P}_k$ indexes the private streams on subcarrier $k$,
$x^{k}_{i,m_i}\in\mathbf{X}_i$ is the symbol of stream $i$ on subcarrier $k$,
$Kp_0+\sum_{i=1}^{J}p_i=P_{\mathrm{tot}}$, and $v_{j,k}\sim\mathcal{CN}(0,\sigma^2)$.

The superimposed constellation on subcarrier $k$ is
\begin{equation}
\mathcal{S}_k=\Big\{\sqrt{p_0}\,x^{k}_{0,m_0}
+\textstyle\sum_{i\in\mathcal{P}_k}\sqrt{p_i}\,x^{k}_{i,m_i}\Big\}.
\label{eq:rsma_superimposed}
\end{equation}

\textit{Remark 1:} If all streams use the same unrotated QAM, different symbol
combinations can map to the same point in $\mathcal{S}_k$. For
example, two private streams with equal power give the same sum when
their symbols are swapped, and such points cannot be separated at any
SNR. The user-wise angles $\phi_j$ are chosen such that all superimposed points
are distinct. The rotation also enlarges the minimum distance of
$\mathcal{S}_k$, which improves the error performance.

User $j$ first detects the common stream jointly over all subcarriers from
$\mathbf{y}_j=[y_{j,1},\ldots,y_{j,K}]^{T}$, by averaging over the private streams. This gives the
posterior $P(x_{0,m_0}\mid\mathbf{y}_j)$, from which the common symbol is decided as
$\hat{m}_0=\arg\max_{m_0}P(x_{0,m_0}\mid\mathbf{y}_j)$. The common stream is then removed from each
subcarrier by soft SIC as
\begin{equation}
\tilde{y}_{j,k}=y_{j,k}-h_{j,k}\sqrt{p_0}\sum_{m_0=1}^{M_0}P(x_{0,m_0}\mid\mathbf{y}_j)\,x^{k}_{0,m_0}.
\label{eq:rsma_sic}
\end{equation}
Next, user $j$ detects its private symbol on its subcarrier $k$, where $\rho_{j,k}=1$. It uses
joint ML detection with the other private streams on that subcarrier, i.e.,
\begin{equation}
\begin{aligned}
\hat{m}_j=\;&\underset{m_j}{\arg\min}\min_{\{m_i\}_{i\in\mathcal{P}_k\setminus\{j\}}}
\Big|\tilde{y}_{j,k}-h_{j,k}\Big(\sqrt{p_j}\,x^{k}_{j,m_j}\\
&+\textstyle\sum_{i\in\mathcal{P}_k\setminus\{j\}}\sqrt{p_i}\,x^{k}_{i,m_i}\Big)\Big|^2.
\end{aligned}
\label{eq:rsma_private_ml}
\end{equation}

\textit{Remark 2:} Conventional RSMA receivers decode the common stream by treating the
private streams as noise, remove it by hard SIC, and decode each
private stream by treating the other private streams as noise. The
considered receiver differs in three ways. First, the common symbol is
detected jointly over all $K$ subcarriers, which gives a diversity
order of $K$ for the common stream. Second, the soft SIC in
(\ref{eq:rsma_sic}) reduces error propagation into private detection.
Third, the joint ML detection in (\ref{eq:rsma_private_ml}) decodes the
co-channel private streams instead of treating them as noise. Hence,
the error floor at high SNR is avoided. 
\subsection{Rate Analysis under Finite Alphabet}
The sum rate of MC-RSMA is the sum of the common and private rates over all subcarriers. The
common stream must be decoded by all users, so its rate is limited by the weakest user. The
achievable rate~\cite{10694077} of the common stream at user $j$ on subcarrier $k$ is given in
(\ref{mc_rsma_common}). After SIC, the rate of the private stream of user $j\in\mathcal{P}_k$ on
subcarrier $k$ is given in (\ref{mc_rsma_private}).
\begin{figure*}[!b]
\hrulefill
\begin{equation}
\resizebox{0.95\textwidth}{!}{$
\begin{aligned}
R_{c,j}^{(k)}
&= \log_2 |M_0| - \frac{J}{\ln 2}
- \frac{1}{|M_0 \times M_{\mathcal{P}_k}|}
\sum_{m=1}^{|M_0 \times M_{\mathcal{P}_k}|}
\mathbb{E}_{v}\Bigg[\log_2
\sum_{l=1}^{|M_0 \times M_{\mathcal{P}_k}|}
\exp\left(-\frac{|\Delta_{m,l}^{(c,k)} + v_j^{(k)}|^2}{\sigma^2}\right)\Bigg]
+ \frac{1}{|M_{\mathcal{P}_k}|}
\sum_{m=1}^{|M_{\mathcal{P}_k}|}
\mathbb{E}_{v}\Bigg[\log_2
\sum_{l=1}^{|M_{\mathcal{P}_k}|}
\exp\left(-\frac{|\Delta_{m,l}^{(p,k)} + v_j^{(k)}|^2}{\sigma^2}\right)\Bigg]
\end{aligned}
$}
\label{mc_rsma_common}
\end{equation}
\begin{equation}
\resizebox{0.95\textwidth}{!}{$
\begin{aligned}
&\approx \log_2 |M_0|
- \frac{1}{|M_0 \times M_{\mathcal{P}_k}|}
\sum_{m=1}^{|M_0 \times M_{\mathcal{P}_k}|}
\log_2
\sum_{l=1}^{|M_0 \times M_{\mathcal{P}_k}|}
\exp\left(-\frac{1}{2\sigma^2}\left|\Delta_{m,l}^{(c,k)}\right|^2\right)
+ \frac{1}{|M_{\mathcal{P}_k}|}
\sum_{m=1}^{|M_{\mathcal{P}_k}|}
\log_2
\sum_{l=1}^{|M_{\mathcal{P}_k}|}
\exp\left(-\frac{1}{2\sigma^2}\left|\Delta_{m,l}^{(p,k)}\right|^2\right)
\end{aligned}
$}
\label{mc_rsma_common_approx}
\end{equation}
\end{figure*}
\begin{figure*}[!b]
\hrulefill
\begin{equation}
\resizebox{0.95\textwidth}{!}{$
\begin{aligned}
R_{p,j}^{(k)}
&= \log_2 |M_j|
- \frac{1}{|M_{\mathcal{P}_k}|}
\sum_{m=1}^{|M_{\mathcal{P}_k}|}
\mathbb{E}_{v}\Bigg[\log_2
\sum_{l=1}^{|M_{\mathcal{P}_k}|}
\exp\left(-\frac{1}{\sigma^2}\left|\Delta_{m,l}^{(p,k)} + v_j^{(k)}\right|^2\right)\Bigg]
+ \frac{1}{|M_{\mathcal{P}_k \setminus \{j\}}|}
\sum_{m=1}^{|M_{\mathcal{P}_k \setminus \{j\}}|}
\mathbb{E}_{v}\Bigg[\log_2
\sum_{l=1}^{|M_{\mathcal{P}_k \setminus \{j\}}|}
\exp\left(-\frac{1}{\sigma^2}\left|\Delta_{m,l}^{(-j,k)} + v_j^{(k)}\right|^2\right)\Bigg]
\end{aligned}
$}
\label{mc_rsma_private}
\end{equation}
\begin{equation}
\resizebox{0.95\textwidth}{!}{$
\begin{aligned}
&\approx \log_2 |M_j|
- \frac{1}{|M_{\mathcal{P}_k}|}
\sum_{m=1}^{|M_{\mathcal{P}_k}|}
\log_2
\sum_{l=1}^{|M_{\mathcal{P}_k}|}
\exp\left(-\frac{1}{2\sigma^2}\left|\Delta_{m,l}^{(p,k)}\right|^2\right)
+ \frac{1}{|M_{\mathcal{P}_k \setminus \{j\}}|}
\sum_{m=1}^{|M_{\mathcal{P}_k \setminus \{j\}}|}
\log_2
\sum_{l=1}^{|M_{\mathcal{P}_k \setminus \{j\}}|}
\exp\left(-\frac{1}{2\sigma^2}\left|\Delta_{m,l}^{(-j,k)}\right|^2\right)
\end{aligned}
$}
\label{mc_rsma_private_approx}
\end{equation}
\end{figure*}
In (\ref{mc_rsma_common})\textendash(\ref{mc_rsma_private_approx}), $M_0$ and $M_j$ are the
constellation sizes of the common and private streams, and
$M_{\mathcal{P}_k}=\prod_{a\in\mathcal{P}_k}M_a$ is the joint constellation size of the streams
on subcarrier $k$. The terms $\Delta_{m,l}$ are the differences between superimposed
constellation points, scaled by $h_{j,k}$.

The total achievable sum rate of MC-RSMA is
\begin{equation}
R_{\mathrm{sum}}=R_c+\sum_{k=1}^{K}\sum_{j=1}^{J}\rho_{j,k}R_{p,j}^{(k)},\qquad
R_c=\min_{j}\Big\{\sum_{k=1}^{K}R_{c,j}^{(k)}\Big\},
\label{eq:mc_rsma_sumrate}
\end{equation}
where $R_c$ is the common rate over all subcarriers and $R_{p,j}^{(k)}$ is the private rate of
user $j$ on subcarrier $k$.
\vspace{-0.1in}
\subsection{Optimized Power Allocation for MC-RSMA}
For a given subcarrier assignment $\{\rho_{j,k}\}$ and channel, the powers $\mathbf{p}$ and the
rotation angles $\boldsymbol{\phi}=[\phi_0,\phi_1,\ldots,\phi_J]^{T}$ are optimized to maximize the sum rate as
\begin{equation}
\begin{aligned}
\max_{\mathbf{p},\,\boldsymbol{\phi}}\;\; & R_{\mathrm{sum}}(\mathbf{p},\boldsymbol{\phi})\\
\text{s.t.}\;\; & Kp_0+\textstyle\sum_{j=1}^{J}p_j\le P_{\mathrm{tot}},\\
& p_0\ge p_j\ge 0,\quad \forall j=1,\ldots,J,
\end{aligned}
\label{eq:sumrate_optimization}
\end{equation}
where $R_{\mathrm{sum}}$ is given by (\ref{eq:mc_rsma_sumrate}). The constraints limit the total
transmit power, keep all powers non-negative, and give the common stream enough power to be
decoded before the private streams. The complete MC-RSMA procedure, including the subcarrier allocation and the genetic algorithm (GA)-based
optimization, is summarized in Algorithm~\ref{alg:rsma_hungarian_ga}.
\begin{algorithm}[!htbp]
\DontPrintSemicolon
\SetAlgoLined
\SetKwInput{KwInit}{Input}
\SetKwInput{KwOutput}{Output}
\small
\caption{\small MC-RSMA With Hungarian Subcarrier Allocation and GA-Based Optimization of Power and Rotation Angles}
\label{alg:rsma_hungarian_ga}
\KwInit{$J$, $K$, $\{M_j\}$, $P_{\mathrm{tot}}$, $\{h_{j,k}\}$, and $\sigma^2$}
\KwOutput{$\boldsymbol{\phi}^{\star}$, $\mathbf{p}^{\star}$, $\{\rho_{j,k}\}$, and $R_{\mathrm{sum}}$}
\BlankLine
\textbf{1) Message Splitting:}\;
\quad Split the messages of the split users into common and private parts.\;
\textbf{2) Subcarrier Allocation:}\;
\quad (i) Allocate the common stream to all subcarriers via (\ref{eq:common_alloc}).\;
\quad (ii) Form the cost matrix from the channel gains $|h_{j,k}|^2$.\;
\quad (iii) If $J>K$, add $J-K$ dummy subcarriers.\;
\quad (iv) Solve (\ref{eq:hungarian}) using the Hungarian algorithm~\cite{9201435}.\;
\quad (v) Reassign the users on dummy subcarriers to $k^{*}=\arg\max_k|h_{j,k}|^2$.\;
\textbf{3) Rotation and Power Optimization:}\;
\quad Obtain $\boldsymbol{\phi}^{\star}$ and $\mathbf{p}^{\star}$ by solving
(\ref{eq:sumrate_optimization}) using GA.\;
\quad Compute $R_{\mathrm{sum}}$ via (\ref{eq:mc_rsma_sumrate}).\;
\textbf{4) Transmission:}\;
\quad Transmit the streams with $\boldsymbol{\phi}^{\star}$, $\mathbf{p}^{\star}$, and
$\{\rho_{j,k}\}$; the received signal follows (\ref{eq:rsma_receiver}).\;
\textbf{5) Receiver Processing:}\;
\quad (i) Detect the common stream jointly over all subcarriers.\;
\quad (ii) Remove the common stream via soft SIC in (\ref{eq:rsma_sic}).\;
\quad (iii) Detect the private streams via (\ref{eq:rsma_private_ml}).\;
\end{algorithm}

\vspace{-0.15in}
\subsection{Complexity Analysis}
\begin{table*}[!htbp]
\centering
\caption{Comparison of the Considered System Models ($K=4$ resources in all cases).}
\label{table:comparison}
\footnotesize
\setlength{\tabcolsep}{4pt}
\renewcommand{\arraystretch}{1.1}
\begin{tabular}{|l|c|c|c|c|c|c|}
\hline
\multirow{2}{*}{\textbf{Metric}}
& \multicolumn{3}{c|}{\textbf{Case I} ($\eta=4.5$ bits/s/Hz)}
& \multicolumn{3}{c|}{\textbf{Case II} ($\eta=3.75$ bits/s/Hz)} \\
\cline{2-7}
& \textbf{RS-SCMA} & \textbf{SCMA} & \textbf{MC-RSMA}
& \textbf{RS-SCMA} & \textbf{SCMA} & \textbf{MC-RSMA} \\
\hline
Number of streams  & 7 & 6 & 7 & 6 & 5 & 6 \\
\hline
Overloading factor & 175\% & 150\% & 175\% & 150\% & 125\% & 150\% \\
\hline
Modulation order
 & \makecell{4 streams: $M=8$\\ 3 streams: $M=4$}
 & \makecell{6 streams:\\ $M=8$}
 & \makecell{4 streams: $M=8$\\ 3 streams: $M=4$}
 & \makecell{3 streams: $M=8$\\ 3 streams: $M=4$}
 & \makecell{5 streams:\\ $M=8$}
 & \makecell{3 streams: $M=8$\\ 3 streams: $M=4$} \\
\hline
Codebook structure & Heterogeneous & Uniform & --- & Heterogeneous & Uniform & --- \\
\hline
Receiver type
& Het.\ MPA / SIC-Het.\ MPA & MPA & SIC
& Het.\ MPA / SIC-Het.\ MPA & MPA & SIC \\
\hline
\end{tabular}
\end{table*}
The detection complexity of the proposed Type-II RS-SCMA system mainly depends on the heterogeneous MPA.
For the non-SIC receiver, the common and private streams are jointly detected, and the
complexity per MPA iteration is
$C_{\mathrm{RS\text{-}SCMA}}^{\mathrm{NS}}=\mathcal{O}\big(\sum_{k=1}^{K}\prod_{j\in\mathcal{M}_k} M_j\big)$,
where $\mathcal{M}_k$ denotes the set of streams connected to resource $k$. For the SIC-based
receiver, the private streams are detected by a second heterogeneous MPA after the common stream
is removed. Including the $\mathcal{O}(K)$ reconstruction and subtraction of the common signal,
the complexity is
$C_{\mathrm{RS\text{-}SCMA}}^{\mathrm{SIC}}=\mathcal{O}\big(\sum_{k=1}^{K}\prod_{j\in\mathcal{M}_k} M_j+\sum_{k=1}^{K}\prod_{j\in\mathcal{M}_k\setminus\{0\}} M_j+K\big)$.
For conventional SCMA, all users use the same modulation order $M$, and the MPA complexity is
$C_{\mathrm{SCMA}}=\mathcal{O}\big(\sum_{k=1}^{K}M^{|\mathcal{M}_k|}\big)$.
For MC-RSMA, the common stream is detected by marginalizing the private streams on each
subcarrier, and each private stream is detected by joint ML with the other private streams on
its subcarrier, which gives
$C_{\mathrm{MC\text{-}RSMA}}=\mathcal{O}\big(\sum_{k=1}^{K}(M_0+1)\prod_{j\in\mathcal{P}_k} M_j\big)$. The additional complexity of RS-SCMA arises mainly from resources shared by four streams. In MC-RSMA, at most two private streams share a subcarrier, so its complexity is lower, at the
cost of the poorer BER and sum-rate performance.
\section{Simulation Results}
\label{sec:results}
Two system configurations are considered to evaluate the proposed Type-II RS-SCMA framework, Case~I with
$\eta=4.5$~bits/s/Hz and Case~II with $\eta=3.75$~bits/s/Hz. Both the cases use heterogeneous modulation
orders and an irregular factor graph. Table~\ref{table:comparison} summarizes the three compared
systems, which all use $K=4$ resources and the same $\eta$ in each case. Due to rate splitting,
RS-SCMA and MC-RSMA have one more stream than SCMA. RS-SCMA uses $M=4$ and $M=8$ with the
heterogeneous MPA, with or without SIC, while SCMA uses $M=8$ with the MPA, and MC-RSMA uses rotated
QAM with an SIC-based ML receiver. The MATLAB code used to generate the
results in this paper is publicly available: \url{https://github.com/kuntaliitg/RS-SCMA-Type-II-MATLAB-Code}

The Type-II RS-SCMA codebook parameters are obtained by minimizing the BER, as in
(\ref{eq:modulation_optimization})\textendash(\ref{eq:beta_optimization}). For MC-RSMA, the
rotation angles and the power allocation are obtained by maximizing the sum rate, as in
(\ref{eq:sumrate_optimization}). Both optimizations use the genetic algorithm (GA) in the MATLAB
Global Optimization Toolbox. The population size and the maximum number of generations are both
set to $50$.


\vspace{-0.1in}
\subsection{Case~I: $\eta=4.5$~bits/s/Hz}

We consider a Type-II RS-SCMA system with $K=4$ REs, derived from an SCMA system with $J=6$ users and
$M=8$. The first bits of two users form the common stream. Their remaining private bits form
streams~5 and~6, and the four unsplit users form streams~1\textendash4. 
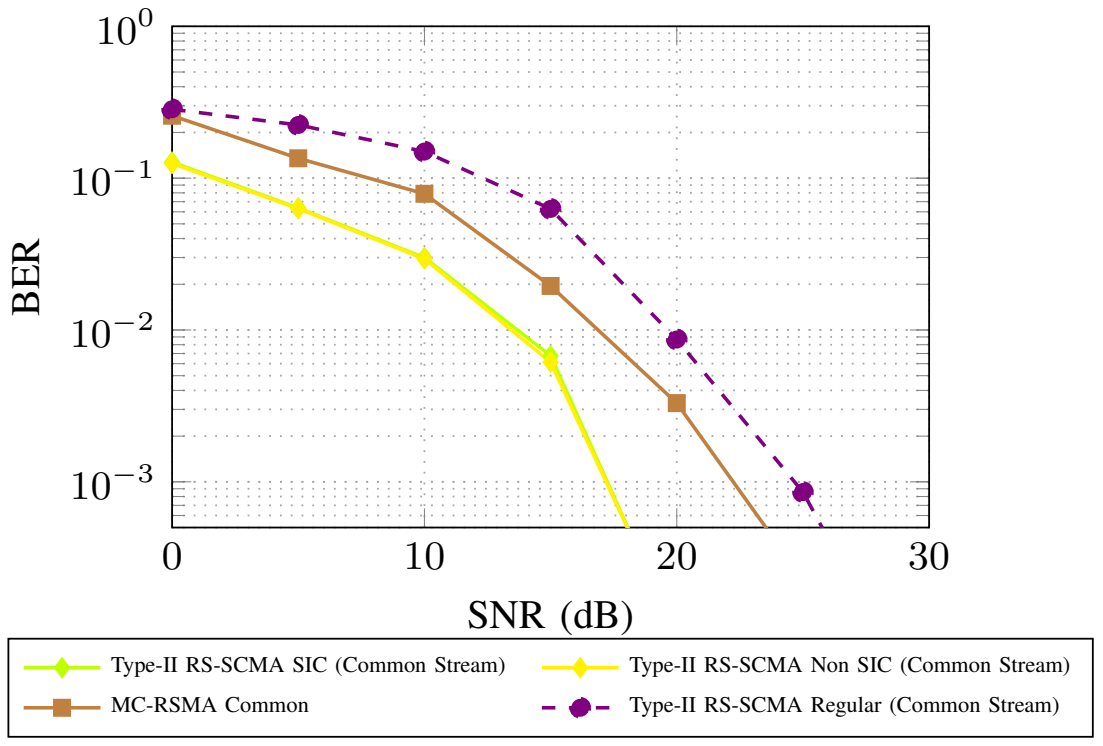
\begin{figure}[!htbp]
\centering
 \resizebox{0.4\textwidth}{!}{%
\begin{tikzpicture}
\begin{semilogyaxis}[
    xlabel={SNR (dB)},
    ylabel={BER},
       xlabel style={font=\small},
    ylabel style={font=\small},
    tick label style={font=\small},
    grid=both,
    grid style={dotted, gray!70},
    width=7.5cm,
    height=5.5cm,
    ymin=5e-4, ymax=1,
    xmin=0, xmax=30,
    legend style={
        at={(0.5,-0.22)},
        anchor=north,
        font=\tiny,
        draw=black,
        /tikz/every even column/.append style={column sep=0.2cm}
    },
    legend cell align=left,
    legend columns=2
]

\addplot[
    lime,
    thick,
    mark=diamond*,
    mark options={scale=0.9}
] coordinates {
(0,0.12745098)
(5,0.063610169)
(10,0.029953917)
(15,0.006756757)
(20,9.58957e-05)
(25,0)
(30,0)
};
\addlegendentry{\tiny Type-II RS-SCMA SIC (Common Stream)}

\addplot[
    yellow,
    thick,
    mark=diamond*,
    mark options={scale=0.9}
]coordinates {
(0,0.125)
(5,0.063008547)
(10,0.029411765)
(15,0.006144995)
(20,9.92654e-05)
(25,0)
(30,0)
};
\addlegendentry{\tiny Type-II RS-SCMA Non SIC (Common Stream)}

\addplot[brown, thick, mark=square*, mark options={scale=0.8}] 
coordinates {
(0,0.2578)
(5,0.1351)
(10,0.0789)
(15,0.0195)
(20,0.0033)
(25,2.2833e-04)
(30,7.5000e-06)
};
\addlegendentry{\tiny MC-RSMA Common}

\addplot[
    color=violet,
    dashed,
    mark=*,
    thick
]
coordinates {
    (0,0.284313725)
    (5,0.224074074)
    (10,0.149700599)
    (15,0.062724014)
    (20,0.008654722)
    (25,0.000856354)
    (30,2.48516E-05)
};
\addlegendentry{\tiny Type-II RS-SCMA Regular (Common Stream)}

\end{semilogyaxis}
\end{tikzpicture}}
\caption{\footnotesize BER vs SNR performance of the common stream for $\eta = 4.5$ bits/s/Hz in Rayleigh channel with $K=4$.}
\label{BER_1_Rayleigh_Common}
\end{figure} 
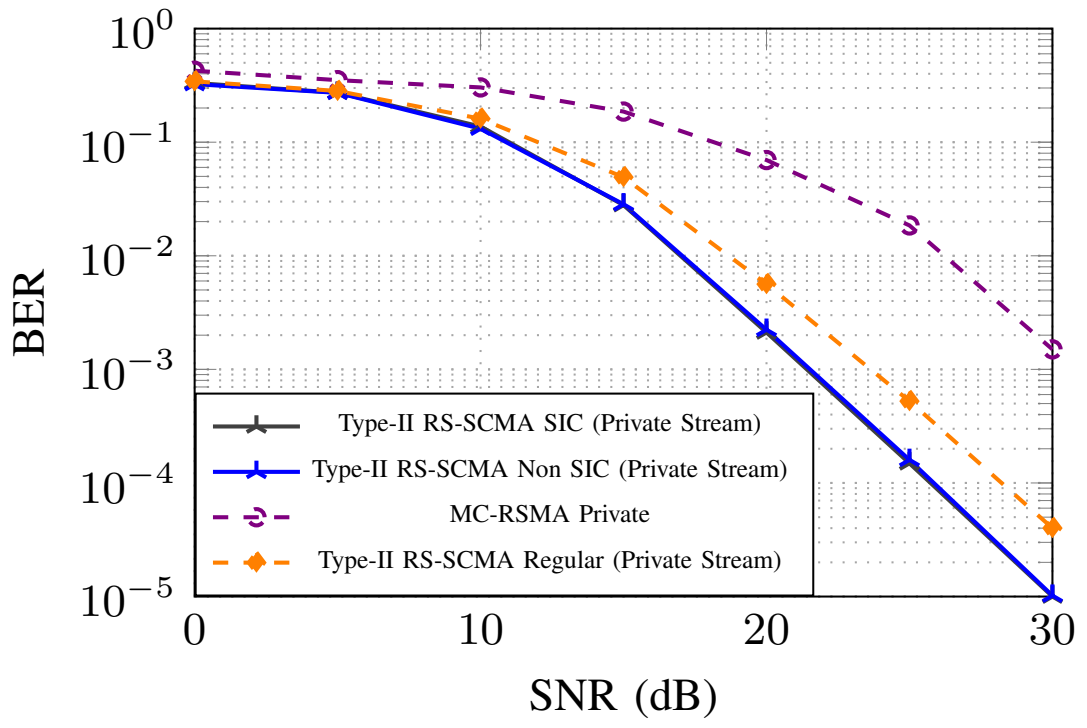
\begin{figure}[!htbp]
\centering
 \resizebox{0.4\textwidth}{!}{%
\begin{tikzpicture}
\begin{semilogyaxis}[
    xlabel={SNR (dB)},
    ylabel={BER},
     xlabel style={font=\small},
    ylabel style={font=\small},
    tick label style={font=\small},
    grid=both,
    grid style={dotted, gray!70},
    width=7.5cm,
    height=5.5cm,
    ymin=1e-5, ymax=1,
    xmin=0, xmax=30,
    legend style={
        at={(0,0)},
        anchor=south west,
        font=\tiny,
        /tikz/every even column/.append style={column sep=0.2cm}
    },
    legend columns=1
]

\addplot[
    darkgray,
    thick,
mark=Mercedes star,
    mark options={scale=1.1}
]coordinates {
(0,0.332107843)
(5,0.272775424)
(10,0.137384793)
(15,0.027927928)
(20,0.002113483)
(25,0.000148869)
(30,1.000e-05)
};
\addlegendentry{\tiny Type-II RS-SCMA SIC (Private Stream)}

\addplot[
    blue,
    thick,
    mark=Mercedes star,
    mark options={scale=1.1}
]coordinates {
(0,0.324519231)
(5,0.271901709)
(10,0.131504525)
(15,0.028414406)
(20,0.002233472)
(25,0.000159709)
(30,1.01944e-05)
};
\addlegendentry{\tiny Type-II RS-SCMA Non SIC (Private Stream)}

\addplot[violet, thick, dashed,  mark=o, mark options={scale=0.8}]coordinates {
(0,0.42575)
(5,0.35175)
(10,0.3035)
(15,0.1860625)
(20,0.069625)
(25,0.0185)
(30,0.0015)
};
\addlegendentry{\tiny MC-RSMA Private}
\addplot[
    color=orange,
    dashed,
    mark=diamond*,
    thick
]
coordinates {
    (0,0.343137255)
    (5,0.282407407)
    (10,0.160928144)
    (15,0.049171147)
    (20,0.005630919)
    (25,0.000524862)
    (30,4.00388E-05)
};
\addlegendentry{Type-II RS-SCMA Regular (Private Stream)}
\end{semilogyaxis}
\end{tikzpicture}}
\caption{ \footnotesize BER vs SNR performance of the private stream for $\eta = 4.5$ bits/s/Hz in Rayleigh channel with $K=4$.}
\label{BER_1_Rayleigh_Private}
\end{figure}
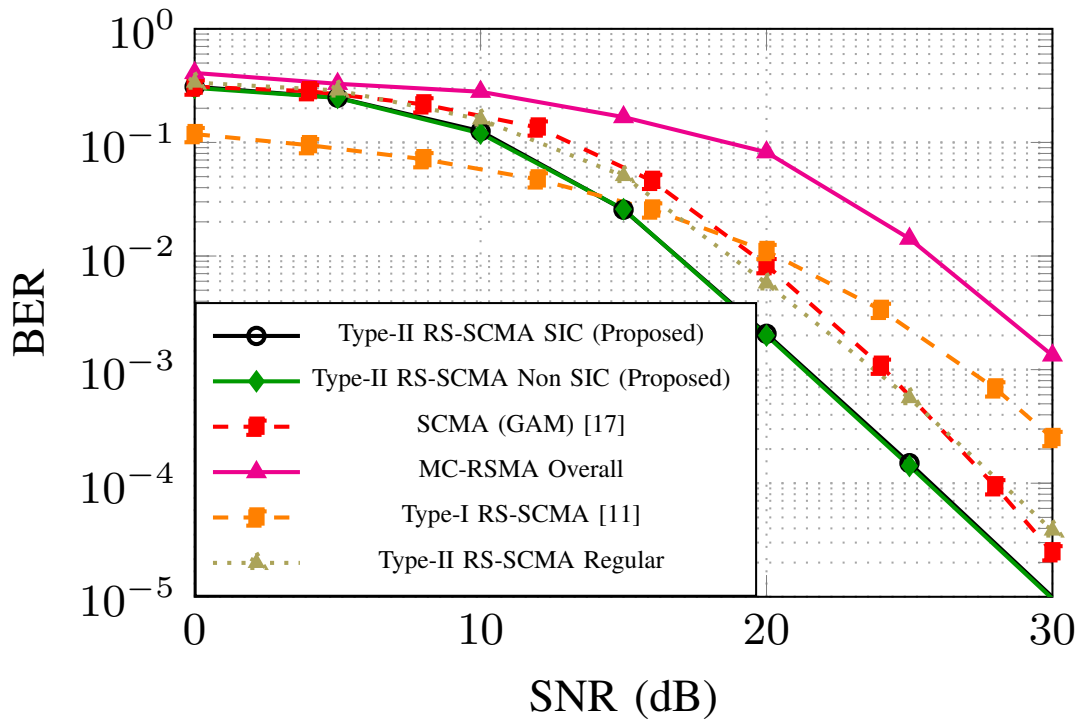
\begin{figure}[!htbp]
\centering
 \resizebox{0.4\textwidth}{!}{%
\begin{tikzpicture}
\begin{semilogyaxis}[
    xlabel={SNR (dB)},
    ylabel={BER},
     xlabel style={font=\small},
    ylabel style={font=\small},
    tick label style={font=\small},
    grid=both,
    grid style={dotted, gray!70},
   width=7.5cm,
    height=5.5cm,
    ymin=1e-5, ymax=1,
    xmin=0, xmax=30,
    legend style={
        at={(0,0)},
        anchor=south west,
        font=\tiny,
        /tikz/every even column/.append style={column sep=0.2cm}
    },
    legend columns=1
]

\addplot[
    black,
    thick,
    mark=o,
    mark options={scale=0.8}
] coordinates {
(0,0.309368192)
(5,0.247645951)
(10,0.125448029)
(15,0.025575576)
(20,0.002067084)
(25,0.000150106)
(30,9.9467e-06)
};
\addlegendentry{\tiny Type-II RS-SCMA SIC (Proposed)}

\addplot[
    green!60!black,
    thick,
  mark=diamond*,
    mark options={scale=0.8}
] coordinates {
(0,0.302350427)
(5,0.24691358)
(10,0.120160885)
(15,0.025828916)
(20,0.001996338)
(25,0.000141963)
(30,9.59502e-06)
};
\addlegendentry{\tiny Type-II RS-SCMA Non SIC (Proposed)}

\addplot[red, thick,dashed,  mark=square*, mark options={scale=0.8}] coordinates {
(0,0.310185185)
(4,0.282622739)
(8,0.216462282)
(12,0.135351298)
(16,0.045902285)
(20,0.008312218)
(24,0.00108223)
(28,9.40014e-05)
(30,2.46586e-05)
};
\addlegendentry{\tiny SCMA (GAM) \cite{8576566}}

\addplot[magenta, thick, mark=triangle*, mark options={scale=0.8}] coordinates {
(0,0.408944444)
(5,0.328333333)
(10,0.2795)
(15,0.167222222)
(20,0.082)
(25,0.014222222)
(30,0.001333333)
};
\addlegendentry{\tiny MC-RSMA Overall}

\addplot[orange, thick,  mark=square*, dashed,mark options={scale=0.8}]
coordinates {
(0, 0.118408333333333)
(4, 0.0946250000000000)
(8, 0.0712000000000000)
(12, 0.0472833333333333)
(16, 0.0256750000000000)
(20, 0.0110750000000000)
(24, 0.00335833333333333)
(28, 0.000683333333333333)
(30, 0.000250000000000000)
};
\addlegendentry{\tiny Type-I RS-SCMA \cite{11493467}}
\addplot[
    color=yellow!60!black,dotted,
    mark=triangle*,
    thick
]
coordinates {
    (0,0.336601307)
    (5,0.287037037)
    (10,0.159680639)
    (15,0.050677021)
    (20,0.005744675)
    (25,0.000561694)
    (30,3.83513E-05)
};
\addlegendentry{Type-II RS-SCMA Regular }
\end{semilogyaxis}
\end{tikzpicture}}
\caption{\footnotesize Overall BER vs SNR performance for $\eta = 4.5$ bits/s/Hz in Rayleigh channel with $K=4$.}
\label{BER_1_Rayleigh_Overall}
\end{figure}
\begin{table}[!htbp]
\centering
\caption{MED and MPD comparison.}
\label{tab:mpd_med}
\footnotesize
\setlength{\tabcolsep}{4pt}
\renewcommand{\arraystretch}{1.1}
\begin{tabular}{|l|c|c|c|}
\hline
\textbf{Case I}
& \textbf{\makecell{RS-SCMA\\(Irregular)}}
& \textbf{\makecell{RS-SCMA\\(Regular)}}
& \textbf{SCMA} \\
\hline
MED & 0.2802 & 0.2251 & 0.2568 \\
\hline
MPD & 0.2911 & 0.2156 & 0.1342 \\
\hline
\end{tabular}
\end{table}
The resulting $J+1=7$
streams have the modulation orders $
\mathbf{M}=[M_0,M_1,\ldots,M_6]=[4,\,8,\,8,\,8,\,8,\,4,\,4],$

where $M_0$ corresponds to the common stream, $M_1,\ldots,M_4$ to the unsplit users, and $M_5,M_6$
to the private streams of the split users. The common stream needs more protection, so the
irregular factor graph
\[
\mathbf{F}_{4\times7}=
\begin{bmatrix}
1 & 1 & 0 & 0 & 1 & 0 & 0\\
1 & 0 & 1 & 0 & 0 & 1 & 1\\
1 & 0 & 0 & 1 & 1 & 1 & 0\\
0 & 1 & 1 & 1 & 0 & 0 & 1
\end{bmatrix}
\]
gives it three REs ($d_{v_0}=3$), while each private stream has two ($d_{v_j}=2$,
$j=1,\ldots,6$).

The codebooks are designed in three steps. The $M=8$ and $M=4$ GAM constellations are generated
with $\theta_0=0.08$ and $0.0635$~\cite{8576566}, and optimized via~\eqref{eq:theta_optimization}
to $\theta^{*}=0.05$ and $0.092$, respectively. The power is then optimized, giving
$\beta^{*}=0.81$, i.e., $0.27P_{\mathrm{tot}}$ per RE for the common stream and about
$0.016P_{\mathrm{tot}}$ per RE for each private stream.

For a fair comparison, all schemes operate at the same spectral efficiency of $\eta=4.5$~bits/s/Hz.
The benchmarks are GAM-SCMA~\cite{8576566} with $J=6$ and $M=8$, Type-I RS-SCMA~\cite{11493467},
and MC-RSMA with $J=6$ users and $K=4$ subcarriers. A Type-II regular RS-SCMA system with
\begin{equation}
\mathbf{F}^{\mathrm{reg}}_{4\times7}=
\begin{bmatrix}
1 & 1 & 0 & 0 & 1 & 0 & 0\\
0 & 0 & 1 & 0 & 0 & 1 & 1\\
1 & 0 & 0 & 1 & 1 & 1 & 0\\
0 & 1 & 1 & 1 & 0 & 0 & 1
\end{bmatrix}.
\label{eq:F_regular}
\end{equation}
is also considered, in which all streams have degree $d_{v_j}=2$. Table~\ref{tab:mpd_med} lists
the MED and MPD. The irregular codebook has the largest MPD ($0.2911$) and MED ($0.2802$). The
regular codebook has a larger MPD than GAM-SCMA ($0.2156$ versus $0.1342$), but a smaller MED
($0.2251$ versus $0.2568$).

For MC-RSMA, the rotation angles $\phi_i$ (in radians) and the powers are optimized at each SNR.
At $20$~dB, they are
\begin{equation}
\begin{aligned}
\boldsymbol{\phi}&=[6.162,\,1.313,\,0.564,\,1.310,\,5.302,\,3.098,\,4.116]^{T},\\
\mathbf{p}&=[0.138,\,0.068,\,0.017,\,0.054,\,0.180,\,0.017,\,0.113]^{T}.
\end{aligned}
\end{equation}
Hence, the common stream gets $Kp_0=0.552P_{\mathrm{tot}}$ in total, which is more than all private
streams together ($0.449P_{\mathrm{tot}}$).

Fig.~\ref{BER_1_Rayleigh_Common} shows the common-stream BER of Case I. The common stream of the proposed
design is the most reliable stream. Its BER drops quickly at medium SNR, and only very few errors
occur at $25$ and $30$~dB. Both receivers give almost the same common BER, since both detect it
with the same MPA over the $K\times(J+1)$ graph. MC-RSMA still has $2.3\times10^{-4}$ at $25$~dB
and $7.5\times10^{-6}$ at $30$~dB. The gain comes from the extra RE of the common stream, which
increases its diversity order.

Fig.~\ref{BER_1_Rayleigh_Private} shows the private-stream BER. The SIC-based receiver gives a
small gain from $15$~dB onwards, e.g., $1.49\times10^{-4}$ versus $1.60\times10^{-4}$ at
$25$~dB, since the common stream is removed before private detection. MC-RSMA has the highest
private BER, still $1.5\times10^{-3}$ at $30$~dB, as each private stream uses a single
subcarrier without spreading diversity.

Fig.~\ref{BER_1_Rayleigh_Overall} compares the overall BER. Both proposed receivers show better BER performance and outperform Type-I RS-SCMA~\cite{11493467} beyond about $15$~dB.
The gap to all benchmarks grows quickly at medium SNR. At $30$~dB, the proposed receivers reach
about $1\times10^{-5}$, versus $2.5\times10^{-5}$, $2.5\times10^{-4}$, and $1.3\times10^{-3}$ for
GAM-SCMA, Type-I RS-SCMA, and MC-RSMA, respectively. This gain comes from the extra RE of the
common stream and from the reliable $M=4$ streams, whose messages help the MPA detect the $M=8$
streams. Both receivers give almost the same overall BER.
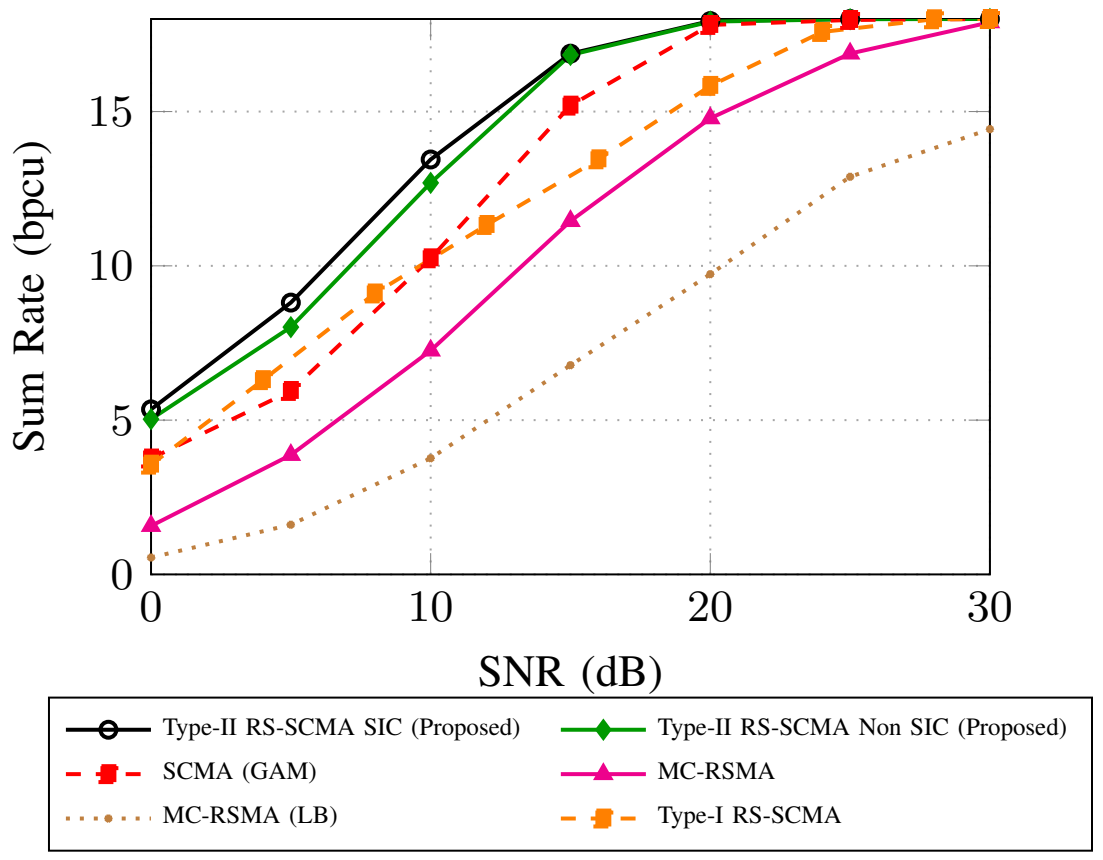
\begin{figure}[!htbp]
\centering
 \resizebox{0.4\textwidth}{!}{%
\begin{tikzpicture}
\begin{axis}[
    xlabel={SNR (dB)},
    ylabel={Sum Rate (bpcu)},
      xlabel style={font=\small},
    ylabel style={font=\small},
    tick label style={font=\small},
    grid=both,
    grid style={dotted, gray!70},
   width=7.5cm,
    height=5.5cm,
    xmin=0, xmax=30,
    ymin=0, ymax=18,
    legend style={
        at={(0.5,-0.22)},
        anchor=north,
        font=\tiny,
        draw=black,
        /tikz/every even column/.append style={column sep=0.2cm}
    },
    legend cell align=left,
    legend columns=2
]
\addplot[
    black,
    thick,
    mark=o,
    mark options={scale=0.8}
]
coordinates {
(0,5.3490)
(5,8.8091)
(10,13.4471)
(15,16.8768)
(20,17.9337)
(25,18.0000)
(30,18.0000)
};
\addlegendentry{\tiny Type-II RS-SCMA SIC (Proposed)}

\addplot[
    green!60!black,
    thick,
    mark=diamond*,
    mark options={scale=0.8}
]coordinates {
(0,5.0316)
(5,8.0138)
(10,12.6891)
(15,16.8353)
(20,17.9150)
(25,17.9938)
(30,18.0000)
};
\addlegendentry{\tiny Type-II RS-SCMA Non SIC (Proposed)}

\addplot[
    red,
    thick,dashed,
    mark=square*,
    mark options={scale=0.8}
]coordinates {
(0,3.7564)
(5,5.9458)
(10,10.2444)
(15,15.1929)
(20,17.8075)
(25,17.9687)
(30,18.0000)
};
\addlegendentry{\tiny SCMA (GAM)}

\addplot[
    magenta,
    thick,
    mark=triangle*,
    mark options={scale=0.8}
] coordinates {
(0,1.570097971)
(5,3.871510564)
(10,7.255147292)
(15,11.45202674)
(20,14.78218492)
(25,16.88045683)
(30,17.89303242)
};
\addlegendentry{\tiny MC-RSMA}

\addplot[
    brown,
    thick,dotted,
    mark=star,
    mark options={scale=0.8}
] coordinates {
(0,0.547020408)
(5,1.61093516)
(10,3.76486545)
(15,6.780509055)
(20,9.729891127)
(25,12.88723086)
(30,14.43045294)
};
\addlegendentry{\tiny MC-RSMA (LB)}

\addplot[orange, thick,dashed, mark=square*, mark options={scale=0.8}]
coordinates {
(0, 3.5578)
(4,6.2961)
(8,9.1053)
(12,11.3248)
(16,13.4526)
(20,15.8324)
(24, 17.5802 )
(28,17.9913)
(30,18)
};
\addlegendentry{\tiny Type-I RS-SCMA}

\end{axis}
\end{tikzpicture}}
\caption{Sum-rate performance versus SNR in Rayleigh fading channel for $\eta = 4.5$ bits/s/Hz.}
\label{sum_1_Rayleigh}
\end{figure}

With $\mathbf{F}^{\mathrm{reg}}_{4\times7}$ (Figs.~\ref{BER_1_Rayleigh_Common}
and~\ref{BER_1_Rayleigh_Private}), the common stream of Type-II RS-SCMA loses this advantage.
From $15$ to $25$~dB, the BER of its common stream exceeds that of its private streams. For
example, at $25$~dB, the common stream of the regular Type-II RS-SCMA has a BER of
$8.6\times10^{-4}$, whereas its private streams have a BER of $5.2\times10^{-4}$. In contrast,
with the irregular graph, the common stream of Type-II RS-SCMA shows very few errors at $25$~dB for both
the SIC-based and non-SIC receivers. By (\ref{eq:pep}), a
stream's error probability decreases with its number of REs. In
$\mathbf{F}^{\mathrm{reg}}_{4\times7}$~\eqref{eq:F_regular}, the common stream has only two REs, both shared with
stream~4 and one shared by four streams, and it is detected in the presence of all private
streams. The irregular graph $\mathbf{F}_{4\times7}$ avoids this by giving it three REs, as
required by (\ref{eq:common_degree}).


Fig.~\ref{sum_1_Rayleigh} shows the sum rate vs SNR for all schemes. At medium SNR, the SIC-based
receiver, as per~\eqref{eq:sum_rate_sic}, achieves the highest sum rate, followed by the non-SIC receiver, as per~\eqref{rate_non_sic}, while GAM-SCMA and Type-I
RS-SCMA perform similarly and MC-RSMA gives the lowest rate. As the SNR increases, the gain of SIC
gradually vanishes, since joint detection is then able to resolve the interference on its own.
At $30$~dB, all schemes except MC-RSMA reach the maximum sum rate of $18$~bpcu, which corresponds
to $\eta=4.5$~bits/s/Hz. MC-RSMA saturates slightly below this at $17.89$~bpcu, whereas its lower
bound reaches only $14.43$~bpcu.

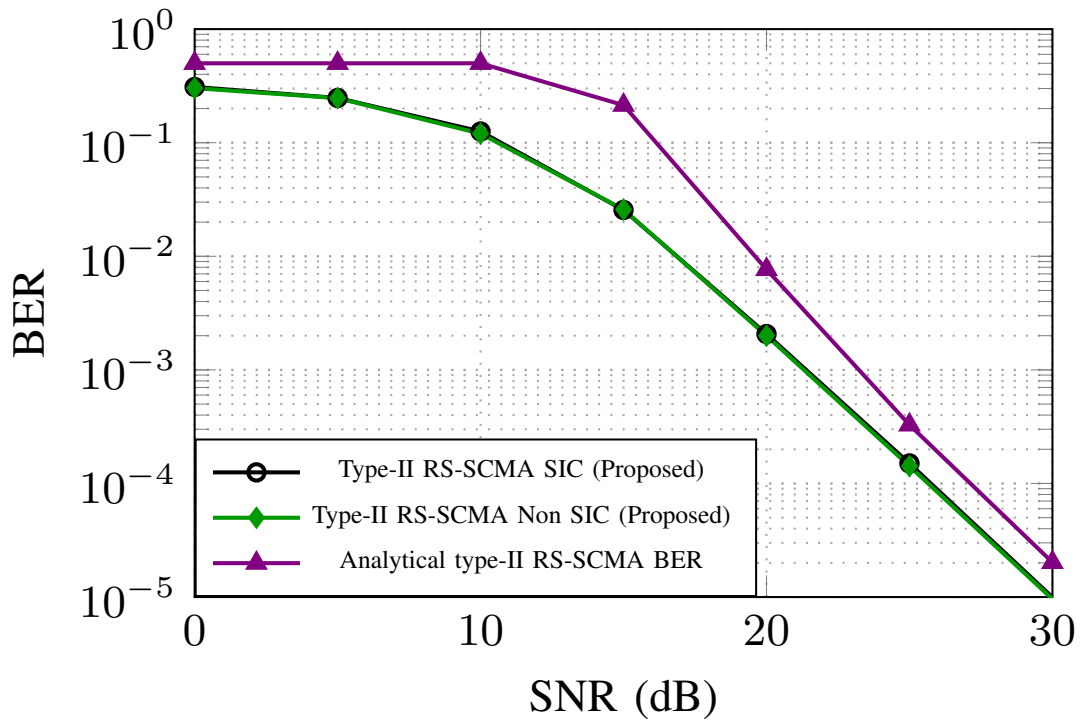
\begin{figure}[!htbp]
\centering
 \resizebox{0.4\textwidth}{!}{%
\begin{tikzpicture}
\begin{semilogyaxis}[
    xlabel={SNR (dB)},
    ylabel={BER},
    xlabel style={font=\small},
    ylabel style={font=\small},
    tick label style={font=\small},
    grid=both,
    grid style={dotted, gray!70},
    width=7.5cm,
    height=5.5cm,
    ymin=1e-5, ymax=1,
    xmin=0, xmax=30,
    legend style={at={(0,0)},anchor=south west},
    legend columns=1
]

\addplot[
    black,
    thick,
    mark=o,
    mark options={scale=0.8}
] coordinates {
(0,0.309368192)
(5,0.247645951)
(10,0.125448029)
(15,0.025575576)
(20,0.002067084)
(25,0.000150106)
(30,9.9467e-06)
};
\addlegendentry{\tiny Type-II RS-SCMA SIC (Proposed)}

\addplot[
    green!60!black,
    thick,
  mark=diamond*,
    mark options={scale=0.8}
] coordinates {
(0,0.302350427)
(5,0.24691358)
(10,0.120160885)
(15,0.025828916)
(20,0.001996338)
(25,0.000141963)
(30,9.59502e-06)
};
\addlegendentry{\tiny Type-II RS-SCMA Non SIC (Proposed)}

\addplot[
    violet,
    thick,
    mark=triangle*,
    mark options={scale=0.9}
]
coordinates {
(0,5.00e-01)
(5,5.00e-01)
(10,5.00e-01)
(15,2.14e-01)
(20,7.66e-03)
(25,3.28e-04)
(30,2.02e-05)
};
\addlegendentry{\tiny Analytical type-II RS-SCMA BER}

\end{semilogyaxis}
\end{tikzpicture}}
\caption{\footnotesize Comparison of simulated vs analytical BER for $\eta = 4.5$ bits/s/Hz in Rayleigh channel with $K=4$ .}
\label{BER_Rayleigh_stsvsana}
\end{figure}

Fig.~\ref{BER_Rayleigh_stsvsana} compares the simulated BER of the proposed Type-II RS-SCMA with the
analytical union bound in (\ref{eq:ber}). At medium SNR, the bound is loose, since overlapping
pairwise error events are counted more than once. At high SNR, the dominant error events take over,
and the bound becomes tight, staying within about twice the simulated BER at $25$ and $30$~dB,
which validates the analysis.

\begin{figure}[!htbp]
\centering
\resizebox{0.4\textwidth}{!}{%
\begin{tikzpicture}
\begin{axis}[
    xlabel={SNR (dB)},
    ylabel={Sum Rate (bpcu)},
    xlabel style={font=\small},
    ylabel style={font=\small},
    tick label style={font=\small},
    grid=both,
    grid style={dotted, gray!70},
   width=7.5cm,
    height=5.5cm,
    xmin=0, xmax=30,
    ymin=0, ymax=20,
    legend style={at={(0,0)},anchor=south west},
    legend columns=2
]

\addplot[
    black,
    thick,
   mark=o,
    mark options={scale=0.8}
]
coordinates {
(0,5.3490)
(5,8.8091)
(10,13.4471)
(15,16.8768)
(20,17.9337)
(25,18.0000)
(30,18.0000)
};
\addlegendentry{\tiny $\epsilon=0$}

\addplot[
    yellow!80!black,
    thick,
    mark=*,
    mark options={scale=0.8}
] coordinates {
    (0,4.44746604007609)
    (5,8.52300893802121)
    (10,11.9311221426519)
    (15,15.3660419682241)
    (20,17.3452658979778)
    (25,17.8210159238602)
    (30,17.9137104517688)
};
\addlegendentry{\tiny $\epsilon=0.05$}

\addplot[
    red,
    thick,
 mark=triangle*,
    mark options={scale=0.9}
] coordinates {
    (0,5.34173511003378)
    (5,7.89039228156928)
    (10,11.0002189735645)
    (15,13.8726350860739)
    (20,16.2318998352554)
    (25,17.6824552589245)
    (30,17.7829625459041)
};
\addlegendentry{\tiny $\epsilon=0.10$}

\addplot[
    magenta,
    thick,dotted,
   mark=triangle*,
    mark options={scale=0.8}
] coordinates {
    (0,4.51273173708130)
    (5,7.29639205877596)
    (10,9.61493495555001)
    (15,11.8089148347194)
    (20,14.4239724138272)
    (25,16.8449104343691)
    (30,16.8830250506758)
};
\addlegendentry{\tiny $\epsilon=0.20$}

\addplot[
    green,
    thick,dashed,
   mark=o,
    mark options={scale=0.8}
]coordinates {
    (0,4.68712487695686)
    (5,6.16511714674047)
    (10,8.05065122071448)
    (15,9.69131009353347)
    (20,11.9629609287597)
    (25,13.8196567390013)
    (30,15.1545039638268)
};
\addlegendentry{\tiny $\epsilon=0.50$}

\addplot[
    blue,
    thick,dashed,
     mark=square*,
    mark options={scale=0.7}
] coordinates {
    (0,4.4171)
    (5,5.6897)
    (10,6.7487)
    (15,8.3114)
    (20,10.3355)
    (25,12.2258)
    (30,13.8060)
};
\addlegendentry{\tiny $\epsilon=1.00$}

\end{axis}
\end{tikzpicture}}
\caption{\footnotesize Sum-rate vs SNR performance for different levels of SIC imperfection in a Rayleigh fading channel with $J+1=7$ streams.}
\label{sumrate_imperfect_SIC_J7_Rayleigh}
\end{figure}
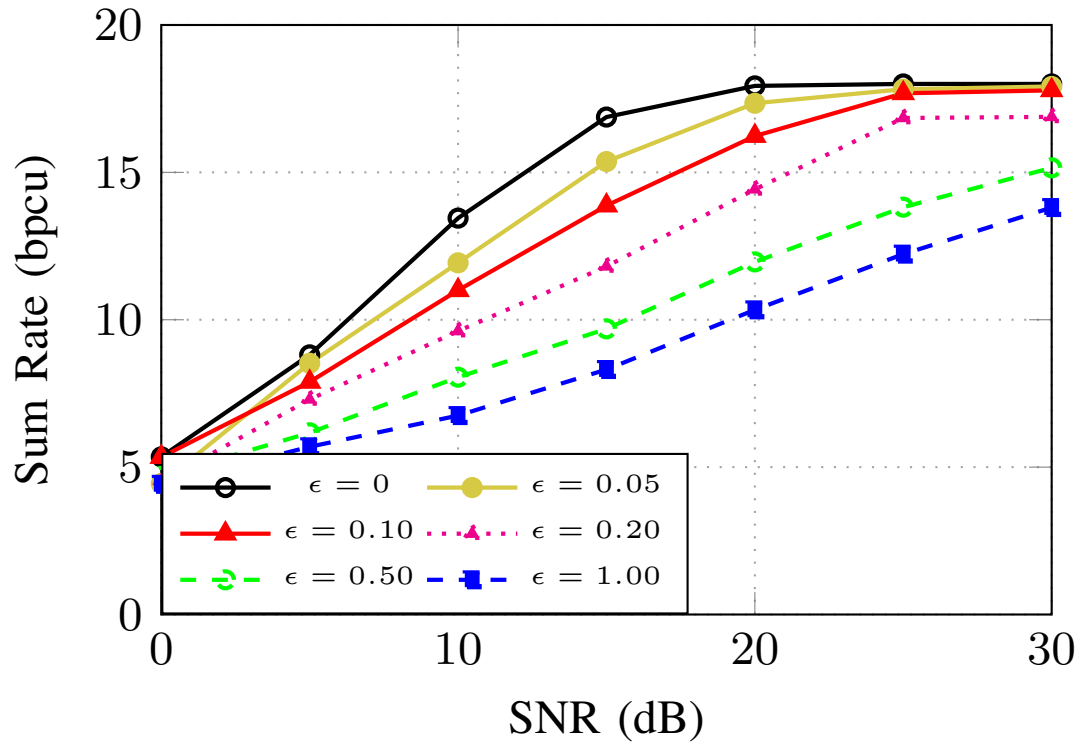
Fig.~\ref{sumrate_imperfect_SIC_J7_Rayleigh} shows the sum rate versus SNR for different levels of
SIC imperfection $\epsilon$, obtained from~\eqref{rate_private_sic}. The curve for $\epsilon=0$
corresponds to perfect SIC and matches the SIC-based curve in Fig.~\ref{sum_1_Rayleigh}. At low SNR,
all curves lie close to each other, since the noise dominates the residual interference. The effect
of $\epsilon$ is most visible at medium SNR. At $15$~dB, for example, the sum rate drops from
$16.88$~bpcu for perfect SIC to $13.87$~bpcu for $\epsilon=0.10$ and to $8.31$~bpcu for
$\epsilon=1.00$. At high SNR, the curves for $\epsilon=0.05$ and $\epsilon=0.10$ still saturate close to the maximum of
$18$~bpcu, reaching $17.91$ and $17.78$~bpcu at $30$~dB, respectively. For $\epsilon=0.20$, the sum rate
saturates at a lower value of $16.88$~bpcu. For $\epsilon=0.50$ and $1.00$, the residual interference
dominates, and the sum rate grows slowly, reaching only $15.15$ and $13.81$~bpcu at $30$~dB. Hence,
the system is robust to mild SIC imperfections, but accurate cancellation is required to achieve the
full sum rate.
\subsection{Case~II: $\eta=3.75$~bits/s/Hz}
The second Type-II RS-SCMA system also uses $K=4$ REs. It is derived from a GAM-SCMA system with $J=5$
users and $M=8$. The first bits of two users are split to form the common stream. The remaining
private bits of these two users form streams~4 and 5. The three unsplit users are mapped to
streams~1\textendash3. Thus, to maintain $\eta=3.75$bits/s/H, the resulting system has $J+1=6$ streams with
$
\mathbf{M}=[M_0,M_1,\ldots,M_5]=[4,\,8,\,8,\,8,\,4,\,4].
$
The irregular factor graph is
{\[
\mathbf{F}_{4\times6}
=
\begin{bmatrix}
1 & 1 & 0 & 0 & 1 & 0\\
1 & 0 & 1 & 0 & 0 & 1\\
1 & 0 & 0 & 1 & 1 & 1\\
0 & 1 & 1 & 1 & 0 & 0
\end{bmatrix}.
\]}
In this graph, the common stream has three REs. Each private stream has two REs.

The codebooks are designed as in Case~I, starting from the same initial GAM parameters, which gives
$\theta^{*}=0.0635$ and $0.09$ for the $M=8$ and $M=4$ codebooks, respectively, and
$\beta^{*}=0.75$. Hence, the common stream gets $0.25P_{\mathrm{tot}}$ on each of its three REs,
and each private stream gets $(1-\beta^{*})P_{\mathrm{tot}}/J=0.05P_{\mathrm{tot}}$, i.e.,
$0.025P_{\mathrm{tot}}$ per RE. All systems operate at $\eta=3.75$~bits/s/Hz. The benchmark
GAM-SCMA~\cite{8576566} uses $J=5$ users with $M=8$ ($5\times3/4=3.75$~bits/s/Hz) and the factor
graph
\[
\mathbf{F}_{4\times5}=
\begin{bmatrix}
1 & 0 & 0 & 1 & 0\\
0 & 1 & 0 & 0 & 1\\
0 & 0 & 1 & 1 & 1\\
1 & 1 & 1 & 0 & 0
\end{bmatrix}.
\]
It is also compared with an MC-RSMA system with $J=5$ users and $K=4$ subcarriers at the same
$\eta=3.75$~bits/s/Hz. For MC-RSMA,
the optimized rotation angles $\boldsymbol{\phi}=[\phi_0,\ldots,\phi_5]$ (in radians) at
$20$~dB are
\begin{equation}
\begin{aligned}
\boldsymbol{\phi}=[\,&1.159,\,0.480,\,5.039,\,0.983,\,3.729,\,4.810\,]^{T}.
\end{aligned}
\end{equation}
The optimized power vector is
\begin{equation}
\mathbf{p}=[0.138,\,0.097,\,0.103,\,0.094,\,0.105,\,0.050]^{T}.
\end{equation}
Hence, the common stream is allocated $Kp_0=0.552P_{\mathrm{tot}}$ in total. This is more than
the total power of all private streams ($0.449P_{\mathrm{tot}}$).

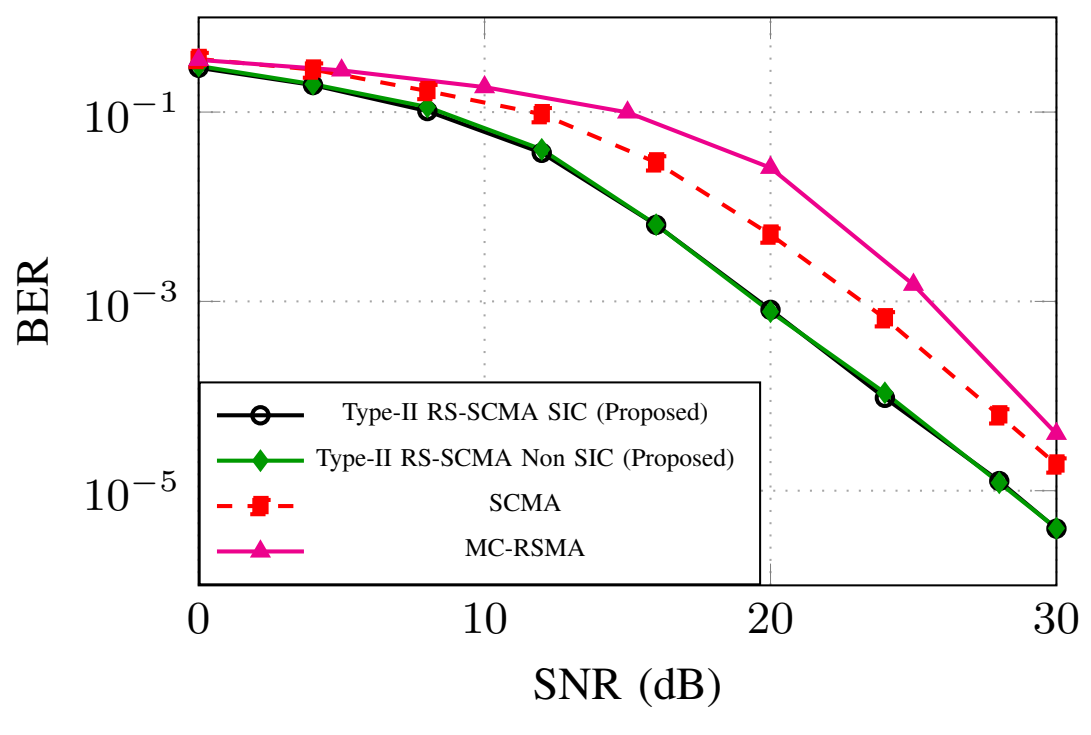
\begin{figure}[!htbp]
\centering
 \resizebox{0.4\textwidth}{!}{%
\begin{tikzpicture}
\begin{semilogyaxis}[
    xlabel={SNR (dB)},
    ylabel={BER},
      xlabel style={font=\small},
    ylabel style={font=\small},
    tick label style={font=\small},
    grid=both,
    grid style={dotted, gray!70},
   width=7.5cm,
    height=5.5cm,
    ymin=1e-6, ymax=1,
    xmin=0, xmax=30,
    legend style={at={(0,0)},anchor=south west}
]
\addplot[black, thick,  mark=o, mark options={scale=0.8}] coordinates {
(0,0.293710691823899) 
(4,0.192982456140351)
(8,0.102459016393443) 
(12,0.0370994940978078) 
(16,0.00643456279443870) 
(20,0.000813874968902157)
(24,9.57314050293177e-05)
(28,1.26420901151169e-05) 
(30,3.97310661755786e-06)
};
\addlegendentry{\tiny Type-II RS-SCMA SIC (Proposed)}
\addplot[green!60!black, thick, mark=diamond*, mark options={scale=0.8}] coordinates {
(0,0.307051282051282) 
(4,0.196875000000000) 
(8,0.113178294573643)
(12,0.0403290414878398)
(16,0.00645475581536963) 
(20,0.000777878977719949)
(24,0.000108105126720168) 
(28,1.21315107020001e-05)
(30,4.00916357835469e-06)
};
\addlegendentry{\tiny Type-II RS-SCMA Non SIC (Proposed)}
\addplot[red, thick,dashed, mark=square*, mark options={scale=0.8}] coordinates {
(0,0.364000000000000) 
(4,0.281269841269841)
(8,0.166899766899767) 
(12,0.0950468540829987) 
(16,0.0294225721784777)
(20,0.00507882159925815)
(24,0.000663449080994889) 
(28,6.25378184780661e-05) 
(30,1.89216785313921e-05)
};
\addlegendentry{\tiny SCMA}
\addplot[magenta, thick,  mark=triangle*, mark options={scale=0.8}]
coordinates {
(0,0.3562)
(5,0.2757)
(10,0.1835)
(15,0.0992)
(20,0.0258)
(25,0.0015)
(30,3.99333e-05)
};
\addlegendentry{\tiny MC-RSMA}
\end{semilogyaxis}
\end{tikzpicture}}
\caption{\footnotesize BER vs SNR performance for $\eta = 3.75$ bits/s/Hz in Rayleigh channel with the number of resources $K=4$.}
\label{BER_2_Rayleigh}
\end{figure}
\vspace{-0.1in}
\begin{figure}[!htbp]
\centering
 \resizebox{0.4\textwidth}{!}{%
\begin{tikzpicture}
\begin{axis}[
    xlabel={SNR (dB)},
    ylabel={Sum Rate (bpcu)},
     xlabel style={font=\small},
    ylabel style={font=\small},
    tick label style={font=\small},
    grid=both,
    grid style={dotted, gray!70},
   width=7.5cm,
    height=5.5cm,
    xmin=0, xmax=30,
    ymin=0, ymax=15,
    legend style={
        at={(0.5,-0.22)},
        anchor=north,
        legend cell align=left,
        font=\tiny,
        draw=black
    },
    legend columns=2
]
\addplot[green!60!black, thick, mark=diamond*, mark options={scale=0.8}] coordinates {
(0,5.0687) 
(4,7.2624) 
(8,9.1559)
(12,11.7797) 
(16,13.5554) 
(20,14.4922) 
(24,14.8163) 
(28,14.9809) 
(30,14.9889)
};
\addlegendentry{\tiny Type-II RS-SCMA Non SIC (Proposed)}
\addplot[black, thick, mark=o, mark options={scale=0.8}] coordinates {
(0,5.7697) 
(4,8.0235) 
(8,10.2643) 
(12,12.2747) 
(16,13.8470) 
(20,14.5911) 
(24,14.8932)
(28,14.9868)
 (30,14.9957)
};
\addlegendentry{\tiny Type-II RS-SCMA SIC (Proposed)}
\addplot[red, thick,dashed, mark=square*, mark options={scale=0.8}] coordinates {
(0,5.2988) 
(4,7.2709) 
(8,9.1347) 
(12,11.7303) 
(16,13.4346) 
(20,14.5181)
(24,14.8478)
(28,14.9803) 
(30,14.9886)
};
\addlegendentry{\tiny SCMA}
\addplot[magenta, thick, mark=triangle*, mark options={scale=0.8}] coordinates {
(0,1.404563406)
(5,3.905558648)
(10,7.122856042)
(15,10.95482101)
(20,14.17124344)
(25,14.98491836)
(30,15)
};
\addlegendentry{\tiny MC-RSMA}
\addplot[brown, thick,dotted, mark=star, mark options={scale=0.8}] coordinates {
(0,0.353090699)
(5,1.746177418)
(10,3.803926929)
(15,6.999890754)
(20,10.5651303)
(25,12.21895417)
(30,12.34347521)
};
\addlegendentry{\tiny MC-RSMA (LB)}
\end{axis}
\end{tikzpicture}}
\caption{Sum-rate performance versus SNR in Rayleigh fading channel for $\eta = 3.75$ bits/s/Hz.}
\label{sum_2_Rayleigh}
\end{figure}
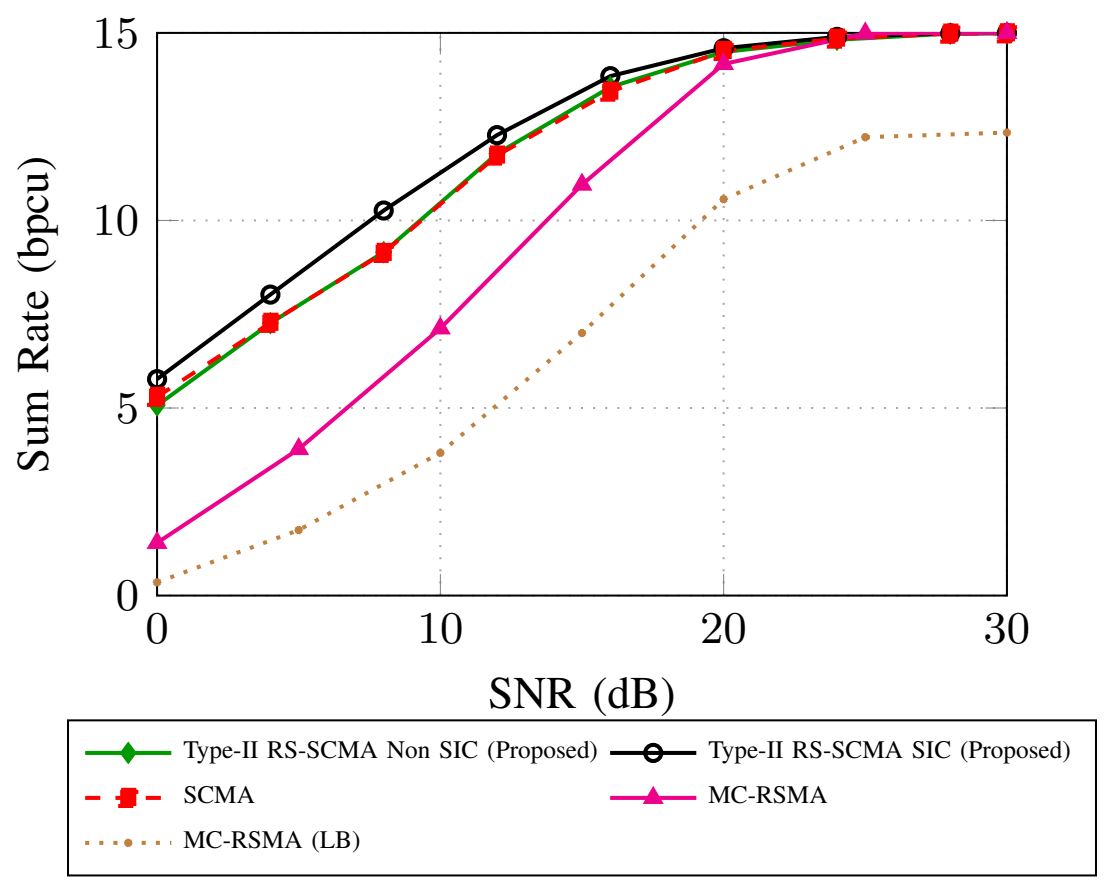
Fig.~\ref{BER_2_Rayleigh} shows the BER of Case~II over the Rayleigh fading channel. Both
proposed receivers outperform GAM-SCMA and MC-RSMA. At medium SNR, the gap grows quickly with
SNR. At a BER of $10^{-4}$, the proposed receivers need about $24$~dB. GAM-SCMA needs about
$27$~dB. This is a gain of about $3$~dB. At $30$~dB, the proposed receivers reach about
$4.0\times10^{-6}$. GAM-SCMA and MC-RSMA reach $1.9\times10^{-5}$ and $4.0\times10^{-5}$,
respectively. The common stream has one more RE than the
private streams. The $M=4$ streams are detected more reliably. They then help the MPA to detect
the $M=8$ streams. The SIC and non-SIC receivers perform similarly. 

The codebooks are designed in three steps. First, the $M=8$ and $M=4$ GAM constellations are
generated with $\theta_0=0.08$ and $0.0635$~\cite{8576566} under unit power. Next, the GAM
parameters are optimized via~\eqref{eq:theta_optimization}, giving $\theta^{*}=0.05$ and $0.092$,
respectively. Finally, the power is optimized with these values fixed, giving $\beta^{*}=0.81$.
Hence, the common stream gets $0.27P_{\mathrm{tot}}$ on each of its three REs, while, by
(\ref{eq:power_split}), each private stream gets about $0.032P_{\mathrm{tot}}$, i.e.,
$0.016P_{\mathrm{tot}}$ per RE.
\subsection{Performance of LDPC-Coded RS-SCMA for Case I }

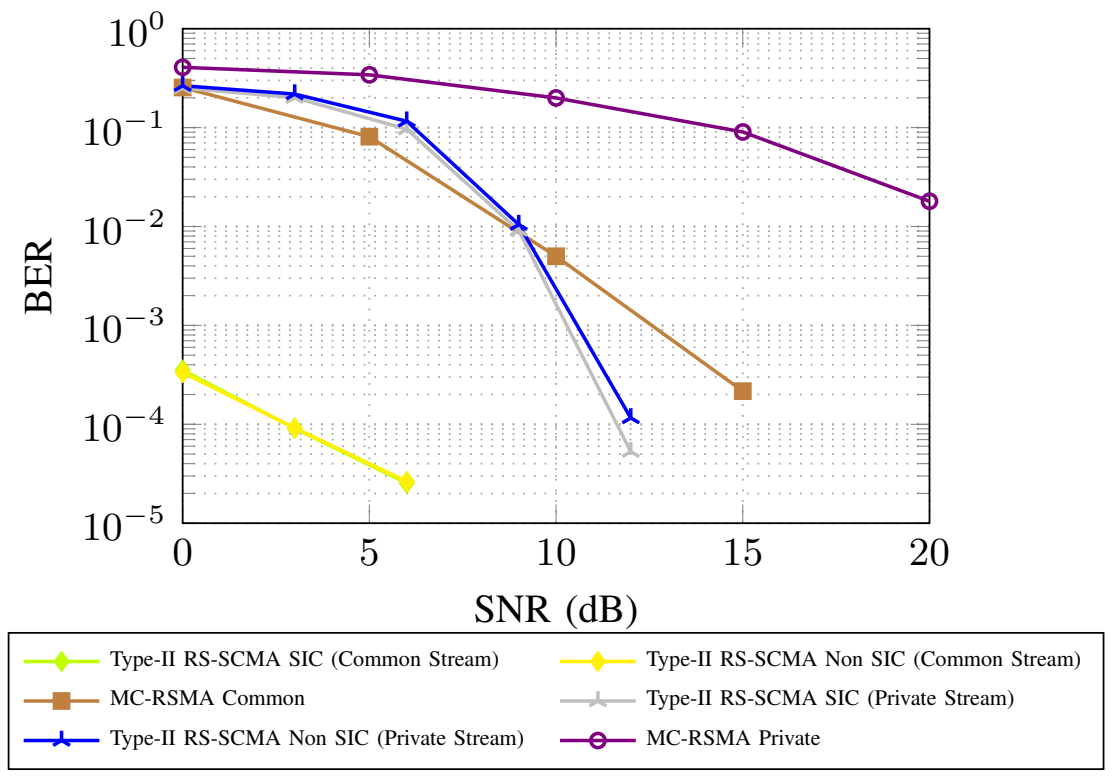
\begin{figure}[!htbp]
\centering
 \resizebox{0.4\textwidth}{!}{%
\begin{tikzpicture}
\begin{semilogyaxis}[
    xlabel={SNR (dB)},
    ylabel={BER},
     xlabel style={font=\small},
    ylabel style={font=\small},
    tick label style={font=\small},
    grid=both,
    grid style={dotted, gray!70},
    width=7.5cm,
    height=5.5cm,
    ymin=1e-5, ymax=1,
    xmin=0, xmax=20,
    legend style={
        at={(0.5,-0.22)},
        anchor=north,
        font=\tiny,
        draw=black,
        /tikz/every even column/.append style={column sep=0.2cm}
    },
    legend cell align=left,
    legend columns=2
]
\addplot[
    lime,
    thick,
    mark=diamond*,
    mark options={scale=0.9}
]
coordinates {
(0,0.000353333)
(3,9.08333e-05)
(6,2.54141e-05)
(9,0)
(12,0)
};
\addlegendentry{\tiny Type-II RS-SCMA SIC (Common Stream)}

\addplot[
    yellow,
    thick,
    mark=diamond*,
    mark options={scale=0.9}
]
coordinates {
(0,0.000333533)
(3,9.18534e-05)
(6,2.64151e-05)
(9,0)
(12,0)
};
\addlegendentry{\tiny Type-II RS-SCMA Non SIC (Common Stream)}

\addplot[brown, thick, mark=square*, mark options={scale=0.8}]
coordinates {
    (0,0.2545)
    (5,0.0809)
    (10,0.0050)
    (15,2.1639e-04)
    (20,0)
};
\addlegendentry{\tiny MC-RSMA Common}

\addplot[
    lightgray,
    thick,
    mark=Mercedes star,
    mark options={scale=1.1}
]
coordinates {
(0,0.251489861)
(3,0.200745833)
(6,0.096401888)
(9,0.009114959)
(12,5.28632e-05)
};
\addlegendentry{\tiny Type-II RS-SCMA SIC (Private Stream)}

\addplot[
    blue,
    thick,
    mark=Mercedes star,
    mark options={scale=1.1}
]
coordinates {
(0,0.2651)
(3,0.218288889)
(6,0.116020833)
(9,0.010449735)
(12,0.000115898)
};
\addlegendentry{\tiny Type-II RS-SCMA Non SIC (Private Stream)}

\addplot[violet, thick, mark=o, mark options={scale=0.8}]
coordinates {
    (0,0.4070)
    (5,0.3416)
    (10,0.1996)
    (15,0.0904)
    (20,0.0180)
};
\addlegendentry{\tiny MC-RSMA Private}

\end{semilogyaxis}
\end{tikzpicture}}
\vspace{0.3cm}
\caption{\footnotesize BER performance versus SNR of the common and private streams for the LDPC-coded systems over the Rayleigh fading channel.}
\label{fig:ber_rayleigh_private}
\end{figure}

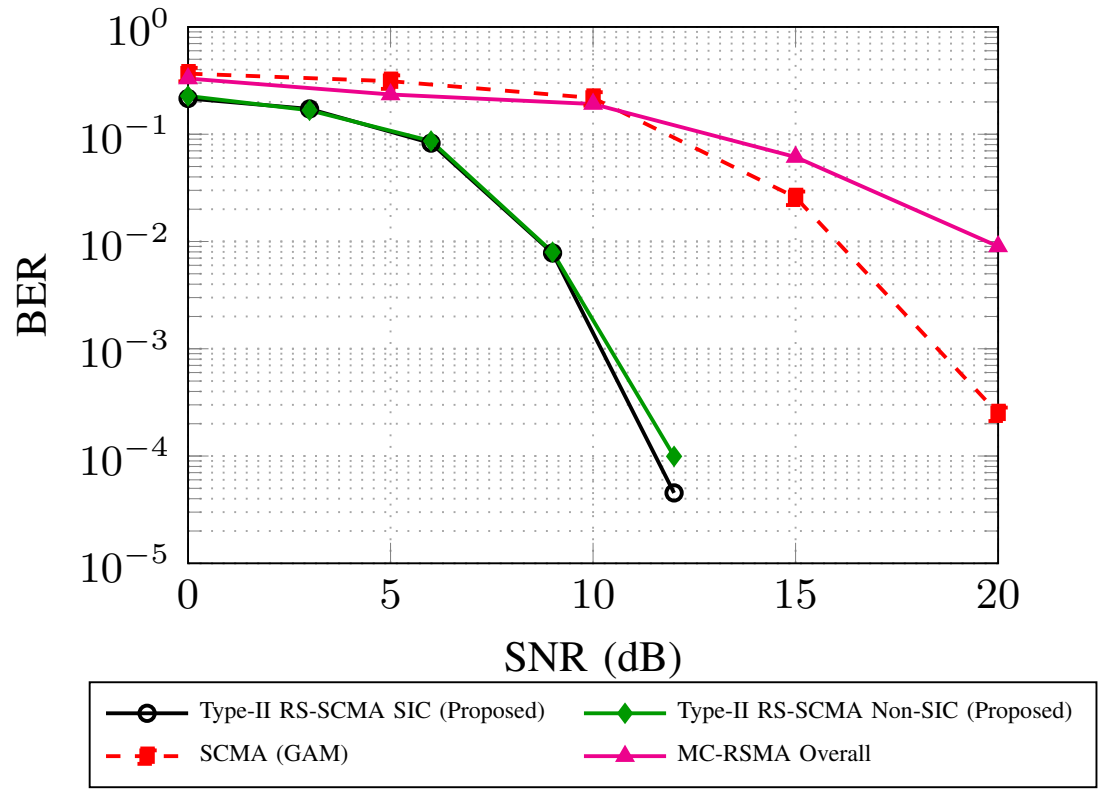
\begin{figure}[!htbp]
\centering
 \resizebox{0.4\textwidth}{!}{%
\begin{tikzpicture}
\begin{semilogyaxis}[
    xlabel={SNR (dB)},
    ylabel={BER},
     xlabel style={font=\small},
    ylabel style={font=\small},
    tick label style={font=\small},
    grid=both,
    grid style={dotted, gray!70},
  width=7.5cm,
    height=5.5cm,
    ymin=1e-5, ymax=1,
    xmin=0, xmax=20,
    legend style={
        at={(0.5,-0.22)},
        anchor=north,
        font=\tiny,
        draw=black,
        /tikz/every even column/.append style={column sep=0.2cm}
    },
    legend cell align=left,
    legend columns=2
]

\addplot[
    black,
    thick,
    mark=o,
    mark options={scale=0.8}
]
coordinates {
(0,0.215613214)
(3,0.17207369)
(6,0.08263382)
(9,0.007812822)
(12,4.53113e-05)
};
\addlegendentry{\tiny Type-II RS-SCMA SIC (Proposed)}

\addplot[
    green!60!black,
    thick,
    mark=diamond*,
    mark options={scale=0.8}
]
coordinates {
(0,0.227254762)
(3,0.167104762)
(6,0.086446429)
(9,0.007956916)
(12,9.93407e-05)
};
\addlegendentry{\tiny Type-II RS-SCMA Non-SIC (Proposed)}

\addplot[red, thick,dashed, mark=square*, mark options={scale=0.8}]
coordinates {
    (0,0.3674)
    (5,0.3124)
    (10,0.2178)
    (15,0.0258)
    (20,2.5000e-04)
};
\addlegendentry{\tiny SCMA (GAM)}

\addplot[magenta, thick, mark=triangle*, mark options={scale=0.8}]
coordinates {
    (0,0.3307)
    (5,0.2347)
    (10,0.1913)
    (15,0.0613)
    (20,0.0090)
};
\addlegendentry{\tiny MC-RSMA Overall}

\end{semilogyaxis}
\end{tikzpicture}}
\vspace{0.3cm}
\caption{\footnotesize Overall BER performance versus SNR for the LDPC-coded systems over the Rayleigh fading channel.}
\label{fig:ber_rayleigh_overall}
\end{figure}
We now consider the LDPC-coded Type-II RS-SCMA system with $J+1=7$ streams and $K=4$ REs. The common and
private streams are encoded with 5G-NR LDPC codes~\cite{9580903}. Their code rates are $r_c$ and
$r_p$, respectively. The spectral efficiency of the coded system is
\begin{equation}
\eta_{\mathrm{coded}}
=
\frac{1}{K}\Big(r_c\log_2 M_0+r_p\sum_{j=1}^{J}\log_2 M_j\Big).
\label{eq:coded_se}
\end{equation}
The code-rate configurations are listed in Table~\ref{tab:code_rates}.

\begin{table}[!t]
\centering
\caption{Configurations of the LDPC-Coded Systems ($K=4$, $\eta=4.5$~bits/s/Hz Uncoded)}
\label{tab:code_rates}
\resizebox{\columnwidth}{!}{%
\begin{tabular}{|l|c|c|c|}
\hline
\textbf{System} & $\boldsymbol{r_c}$ & $\boldsymbol{r_p}$ & $\boldsymbol{\eta_{\mathrm{coded}}}$ \textbf{(bits/s/Hz)} \\
\hline
RS-SCMA, Config.~A ($r_c<r_p$) & 0.33 & 0.40 & 1.77 \\
\hline
RS-SCMA, Config.~B ($r_c>r_p$) & 0.40 & 0.33 & 1.53 \\
\hline
RS-SCMA, Config.~C ($r_c=r_p$) & 0.33 & 0.33 & 1.50 \\
\hline
GAM-SCMA~\cite{8576566}        & \multicolumn{2}{c|}{0.30} & 1.35 \\
\hline
MC-RSMA                         & 0.33 & 0.40 & 1.77 \\
\hline
\end{tabular}%
}
\end{table}
Fig. \ref{fig:ber_rayleigh_private} shows the common- and
private-stream BER with Config.~A. With coding, the common stream is highly reliable. Very few
errors occur from $9$~dB onwards. In contrast, MC-RSMA still has a common-stream BER of
$2.2\times10^{-4}$ at $15$~dB. For the private streams, SIC gives a lower BER. The gain is
largest at high SNR. At $12$~dB, the BER is $5.3\times10^{-5}$ with SIC and $1.2\times10^{-4}$
without it. This is because the LDPC decoder output is more reliable than the uncoded MPA
output. MC-RSMA still has a private-stream BER of $1.8\times10^{-2}$ at $20$~dB.

Fig.~\ref{fig:ber_rayleigh_overall} shows the overall BER. Both proposed receivers clearly
outperform GAM-SCMA and MC-RSMA. At $12$~dB, the BER is $4.5\times10^{-5}$ with SIC and
$9.9\times10^{-5}$ without it. Even at $20$~dB, GAM-SCMA and MC-RSMA reach only
$2.5\times10^{-4}$ and $9.0\times10^{-3}$ respectively. MC-RSMA uses the same code rates, and GAM-SCMA uses a
lower code (Table~\ref{tab:code_rates}). Hence, the gain comes from the RS-SCMA structure.

\begin{figure}[!htbp]
\centering
\resizebox{0.4\textwidth}{!}{%
\begin{tikzpicture}
\begin{semilogyaxis}[
    xlabel={SNR (dB)},
    ylabel={BER},
      xlabel style={font=\small},
    ylabel style={font=\small},
    tick label style={font=\small},
    grid=both,
    grid style={dotted, gray!70},
   width=7.5cm,
    height=5.5cm,
    ymin=1e-6, ymax=1,
    xmin=0, xmax=12,
    legend style={
        at={(0.5,-0.25)},
        anchor=north,
        legend cell align=left,
        draw=black
    },
    legend columns=2
]

\addplot[
    blue,
    thick,
    solid,
    mark=triangle*,
    mark options={scale=0.8}
]
coordinates {
    (0,0.211928571)
    (3,0.16497619)
    (6,0.059269533)
    (9,0.002169353)
    (12,1.0432e-06)
};
\addlegendentry{\tiny SIC-MPA, Config.~C ($r_c=0.33$, $r_p=0.33$)}

\addplot[
    red,
    thick,
    mark=square*,
    mark options={scale=0.8}
]
coordinates {
    (0,0.2170)
    (3,0.1824)
    (6,0.0964)
    (9,0.0150)
    (12,2.3810e-04)
};
\addlegendentry{\tiny SIC-MPA, Config.~B ($r_c=0.40$, $r_p=0.33$)}

\addplot[
    black,
    thick,
    mark=o,
    mark options={scale=0.8}
]
coordinates {
(0,0.215613214)
(3,0.17207369)
(6,0.08263382)
(9,0.007812822)
(12,4.53113e-05)
};
\addlegendentry{\tiny SIC-MPA, Config.~A ($r_c=0.33$, $r_p=0.40$)}

\addplot[
    orange,
    thick,
    mark=star,
    mark options={scale=1}
]
coordinates {
    (0,0.2233)
    (3,0.1846)
    (6,0.0788)
    (9,0.0032)
    (12,1.0848e-06)
};
\addlegendentry{\tiny Non-SIC MPA, Config.~C ($r_c=0.33$, $r_p=0.33$)}

\addplot[
    magenta,
    thick,
    mark=square*,
    mark options={scale=0.8}
]
coordinates {
    (0,0.2287)
    (3,0.1950)
    (6,0.1118)
    (9,0.0160)
    (12,2.4770e-04)
};
\addlegendentry{\tiny Non-SIC MPA, Config.~B ($r_c=0.40$, $r_p=0.33$)}

\addplot[
    green!60!black,
    thick,
    mark=diamond*,
    mark options={scale=0.8}
]
coordinates {
(0,0.227254762)
(3,0.167104762)
(6,0.086446429)
(9,0.007956916)
(12,9.93407e-05)
};
\addlegendentry{\tiny Non-SIC MPA, Config.~A ($r_c=0.33$, $r_p=0.40$)}

\end{semilogyaxis}
\end{tikzpicture}}
\caption{\footnotesize Overall BER versus SNR of the LDPC-coded RS-SCMA system ($J+1=7$, $K=4$) for different common and private code-rate configurations with SIC-based and non-SIC heterogeneous MPA receivers.}
\label{fig:ber_overall}
\end{figure}
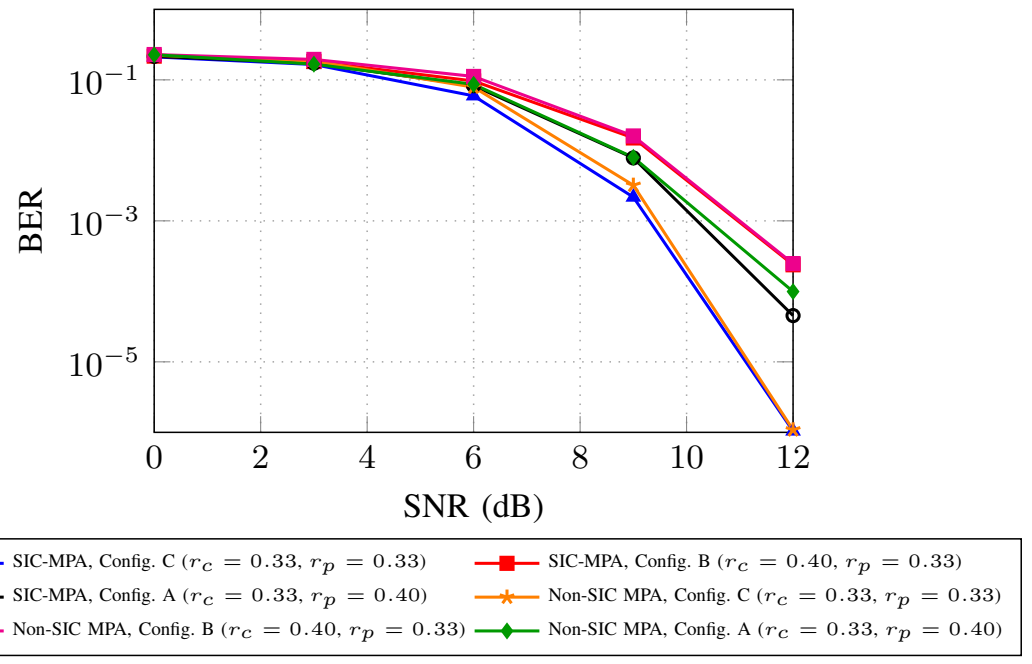

Fig.~\ref{fig:ber_overall} compares the three configurations. Config.~C gives the lowest BER,
$1.0\times10^{-6}$ at $12$~dB with SIC, but has the lowest spectral efficiency. Config.~A
outperforms Config.~B at $12$~dB ($4.5\times10^{-5}$ versus $2.4\times10^{-4}$), although it has
a higher spectral efficiency. This is because a stronger common code makes SIC more reliable.
Without SIC, the gain is smaller ($9.9\times10^{-5}$ versus $2.5\times10^{-4}$).
In summary, coded Type-II RS-SCMA achieves the lowest BER at equal or higher spectral efficiency than
the benchmarks.
\section{Conclusion}


This paper proposed a novel RS-SCMA framework for overloaded downlink systems with heterogeneous modulation orders. By combining rate splitting with SCMA, the proposed scheme improves interference management while maintaining the same spectral efficiency. The unsplit users use higher-order modulation, while the common and private streams use lower-order modulation. A heterogeneous MPA receiver and a GAM-based codebook with optimized rotation and power allocation were developed. For a fair comparison, an MC-RSMA benchmark was also designed with user-wise rotated QAM codebooks, Hungarian-based user-subcarrier matching, and joint ML detection of the private streams. Simulation results for two overloaded systems show that Type-II RS-SCMA achieves better BER and achievable sum rate than conventional SCMA and MC-RSMA at the same spectral efficiency. Type-II RS-SCMA also retains its advantage with LDPC coding. A lower code rate for the common stream gives the best BLER with the SIC-based receiver. These results demonstrate that Type-II RS-SCMA is a promising multiple access technique for future wireless communication systems.

\appendices
\bibliographystyle{IEEEtran} 
\bibliography{references}

\end{document}